\pdfoutput=1

\documentclass[12pt,a4paper]{article}

\usepackage{ifthen} 
\newboolean{pdflatex}
\setboolean{pdflatex}{true} 

\newboolean{articletitles}
\setboolean{articletitles}{true} 

\newboolean{uprightparticles}
\setboolean{uprightparticles}{false} 

\newboolean{inbibliography}
\setboolean{inbibliography}{false} 

\usepackage{microtype}
\usepackage{lineno}  
\usepackage{xspace} 
\usepackage{caption} 

\usepackage{graphicx}  
\usepackage{color}
\usepackage{colortbl}
\graphicspath{{./figs/}} 

\usepackage{amsmath} 
\usepackage{amssymb}
\usepackage{amsfonts}
\usepackage{upgreek} 

\newcommand*\patchAmsMathEnvironmentForLineno[1]{%
\expandafter\let\csname old#1\expandafter\endcsname\csname #1\endcsname
\expandafter\let\csname oldend#1\expandafter\endcsname\csname
end#1\endcsname
 \renewenvironment{#1}%
   {\linenomath\csname old#1\endcsname}%
   {\csname oldend#1\endcsname\endlinenomath}%
}
\newcommand*\patchBothAmsMathEnvironmentsForLineno[1]{%
  \patchAmsMathEnvironmentForLineno{#1}%
  \patchAmsMathEnvironmentForLineno{#1*}%
}
\AtBeginDocument{%
\patchBothAmsMathEnvironmentsForLineno{equation}%
\patchBothAmsMathEnvironmentsForLineno{align}%
\patchBothAmsMathEnvironmentsForLineno{flalign}%
\patchBothAmsMathEnvironmentsForLineno{alignat}%
\patchBothAmsMathEnvironmentsForLineno{gather}%
\patchBothAmsMathEnvironmentsForLineno{multline}%
\patchBothAmsMathEnvironmentsForLineno{eqnarray}%
}

\usepackage{hyperref}    
\usepackage[all]{hypcap} 

\def\lhcb {\mbox{LHCb}\xspace}
\def\ux85 {\mbox{UX85}\xspace}

\def\lhc    {\mbox{LHC}\xspace}

\ifthenelse{\boolean{uprightparticles}}%
{
 
 \def\Pgamma      {\ensuremath{\upgamma}\xspace}

 \def\Pmu         {\ensuremath{\upmu}\xspace}

 \def\Pchi        {\ensuremath{\upchi}\xspace}                 
 \def\Ppsi        {\ensuremath{\uppsi}\xspace}

 \def\PDelta      {\ensuremath{\Delta}\xspace}                 
 \def\PXi      {\ensuremath{\Xi}\xspace}                 
 \def\PLambda      {\ensuremath{\Lambda}\xspace}                 
 \def\PSigma      {\ensuremath{\Sigma}\xspace}                 
 \def\POmega      {\ensuremath{\Omega}\xspace}                 
 \def\PUpsilon      {\ensuremath{\Upsilon}\xspace}                 
 
 \def\PB      {\ensuremath{\mathrm{B}}\xspace}                 
                  
 \def\PD      {\ensuremath{\mathrm{D}}\xspace}

 \def\PJ      {\ensuremath{\mathrm{J}}\xspace}                 
 \def\PK      {\ensuremath{\mathrm{K}}\xspace}

 \def\Pb      {\ensuremath{\mathrm{b}}\xspace}                 
 \def\Pc      {\ensuremath{\mathrm{c}}\xspace}                 
                  
 \def\Pe      {\ensuremath{\mathrm{e}}\xspace}

 \def\Pi      {\ensuremath{\mathrm{i}}\xspace}

 \def\Pq      {\ensuremath{\mathrm{q}}\xspace}

}
{
 
 \def\Pgamma      {\ensuremath{\gamma}\xspace}

 \def\Pmu         {\ensuremath{\mu}\xspace}

 \def\Pchi        {\ensuremath{\chi}\xspace}                 
 \def\Ppsi        {\ensuremath{\psi}\xspace}                 
                  
 \mathchardef\PDelta="7101
 \mathchardef\PXi="7104
 \mathchardef\PLambda="7103
 \mathchardef\PSigma="7106
 \mathchardef\POmega="710A
 \mathchardef\PUpsilon="7107
                  
 \def\PB      {\ensuremath{B}\xspace}                 
                  
 \def\PD      {\ensuremath{D}\xspace}

 \def\PJ      {\ensuremath{J}\xspace}                 
 \def\PK      {\ensuremath{K}\xspace}

 \def\Pb      {\ensuremath{b}\xspace}                 
 \def\Pc      {\ensuremath{c}\xspace}                 
                  
 \def\Pe      {\ensuremath{e}\xspace}

 \def\Pi      {\ensuremath{i}\xspace}

 \def\Pq      {\ensuremath{q}\xspace}

}

\def\en         {\ensuremath{\Pe^-}\xspace}   
\def\ep         {\ensuremath{\Pe^+}\xspace}

\def\mup        {\ensuremath{\Pmu^+}\xspace}
\def\mun        {\ensuremath{\Pmu^-}\xspace} 
\def\mumu       {\ensuremath{\Pmu^+\Pmu^-}\xspace}

\def\g      {\ensuremath{\Pgamma}\xspace}

\def\quark     {\ensuremath{\Pq}\xspace}
\def\quarkbar  {\ensuremath{\overline \quark}\xspace}
\def\qqbar     {\ensuremath{\quark\quarkbar}\xspace}

\def\cquark    {\ensuremath{\Pc}\xspace}
\def\cquarkbar {\ensuremath{\overline \cquark}\xspace}
\def\ccbar     {\ensuremath{\cquark\cquarkbar}\xspace}
\def\bquark    {\ensuremath{\Pb}\xspace}
\def\bquarkbar {\ensuremath{\overline \bquark}\xspace}
\def\bbbar     {\ensuremath{\bquark\bquarkbar}\xspace}

\def\kaon  {\ensuremath{\PK}\xspace}
  \def\Kbar  {\kern 0.2em\overline{\kern -0.2em \PK}{}\xspace}

\def\Kz    {\ensuremath{\kaon^0}\xspace}
\def\Kzb   {\ensuremath{\Kbar^0}\xspace}
\def\KzKzb {\ensuremath{\Kz \kern -0.16em \Kzb}\xspace}
\def\Kp    {\ensuremath{\kaon^+}\xspace}
\def\Km    {\ensuremath{\kaon^-}\xspace}

\def\KpKm  {\ensuremath{\Kp \kern -0.16em \Km}\xspace}

  \def\Dbar    {\kern 0.2em\overline{\kern -0.2em \PD}{}\xspace}
\def\D       {\ensuremath{\PD}\xspace}

\def\Dz      {\ensuremath{\D^0}\xspace}
\def\Dzb     {\ensuremath{\Dbar^0}\xspace}
\def\DzDzb   {\ensuremath{\Dz {\kern -0.16em \Dzb}}\xspace}
\def\Dp      {\ensuremath{\D^+}\xspace}
\def\Dm      {\ensuremath{\D^-}\xspace}

\def\DpDm    {\ensuremath{\Dp {\kern -0.16em \Dm}}\xspace}

\def\B       {\ensuremath{\PB}\xspace}
  \def\Bbar    {\kern 0.18em\overline{\kern -0.18em \PB}{}\xspace}

\def\BBbar   {\ensuremath{\B\Bbar}\xspace}

\def\jpsi     {\ensuremath{{\PJ\mskip -3mu/\mskip -2mu\Ppsi\mskip 2mu}}\xspace}

\def\chicone  {\ensuremath{\Pchi_{\cquark 1}}\xspace}
\def\chictwo  {\ensuremath{\Pchi_{\cquark 2}}\xspace}
\def\chibzero {\ensuremath{\Pchi_{\bquark 0}}}
\def\chibone  {\ensuremath{\Pchi_{\bquark 1}}}
\def\chibtwo  {\ensuremath{\Pchi_{\bquark 2}}}
  \def\Y#1S{\ensuremath{\PUpsilon{(#1S)}}\xspace}
\def\OneS  {\Y1S}
\def\TwoS  {\Y2S}
\def\ThreeS{\Y3S}

\def\chic  {\ensuremath{\Pchi_{\cquark}}}
\def\chib  {\ensuremath{\Pchi_{\bquark}}}

\def\Lbar {\ensuremath{\kern 0.1em\overline{\kern -0.1em\PLambda}}\xspace}

\def\BF         {{\ensuremath{\cal B}\xspace}}

\def\BR         {\BF}

\def\to                 {\ensuremath{\rightarrow}\xspace}

\def\AT#1     {\ensuremath{A_{\mathrm{T}}^{#1}}\xspace}           

\def\C#1      {\ensuremath{\mathcal{C}_{#1}}\xspace}                       
\def\Cp#1     {\ensuremath{\mathcal{C}_{#1}^{'}}\xspace}                    
\def\Ceff#1   {\ensuremath{\mathcal{C}_{#1}^{\mathrm{(eff)}}}\xspace}        
\def\Cpeff#1  {\ensuremath{\mathcal{C}_{#1}^{'\mathrm{(eff)}}}\xspace}       
\def\Ope#1    {\ensuremath{\mathcal{O}_{#1}}\xspace}                       
\def\Opep#1   {\ensuremath{\mathcal{O}_{#1}^{'}}\xspace}                    

\newcommand{\tev}{\ensuremath{\mathrm{\,Te\kern -0.1em V}}\xspace}
\newcommand{\gev}{\ensuremath{\mathrm{\,Ge\kern -0.1em V}}\xspace}
\newcommand{\mev}{\ensuremath{\mathrm{\,Me\kern -0.1em V}}\xspace}
\newcommand{\kev}{\ensuremath{\mathrm{\,ke\kern -0.1em V}}\xspace}
\newcommand{\ev}{\ensuremath{\mathrm{\,e\kern -0.1em V}}\xspace}
\newcommand{\gevc}{\ensuremath{{\mathrm{\,Ge\kern -0.1em V\!/}c}}\xspace}
\newcommand{\gevcnosp}{\ensuremath{{\mathrm{Ge\kern -0.1em V\!/}c}}\xspace}
\newcommand{\mevc}{\ensuremath{{\mathrm{\,Me\kern -0.1em V\!/}c}}\xspace}
\newcommand{\gevcc}{\ensuremath{{\mathrm{\,Ge\kern -0.1em V\!/}c^2}}\xspace}
\newcommand{\gevgevcccc}{\ensuremath{{\mathrm{\,Ge\kern -0.1em V^2\!/}c^4}}\xspace}
\newcommand{\mevcc}{\ensuremath{{\mathrm{\,Me\kern -0.1em V\!/}c^2}}\xspace}

\def\mum  {\ensuremath{\,\upmu\rm m}\xspace}

\def\invfb   {\ensuremath{\mbox{\,fb}^{-1}}\xspace}

\newcommand{\stat}{\ensuremath{\mathrm{(stat)}}\xspace}
\newcommand{\expe}{\ensuremath{\mathrm{(exp)}}\xspace}
\newcommand{\syst}{\ensuremath{\mathrm{(syst)}}\xspace}
\newcommand{\model}{\ensuremath{\mathrm{(model)}}\xspace}

\newcommand{\chisq}{\ensuremath{\chi^2}\xspace}

\def\gsim{{~\raise.15em\hbox{$>$}\kern-.85em
          \lower.35em\hbox{$\sim$}~}\xspace}
\def\lsim{{~\raise.15em\hbox{$<$}\kern-.85em
          \lower.35em\hbox{$\sim$}~}\xspace}

\def\ptot       {\mbox{$p$}\xspace}
\def\pt         {\mbox{$p_{\rm T}$}\xspace}

\def\evtgen     {\mbox{\textsc{EvtGen}}\xspace}
\def\pythia     {\mbox{\textsc{Pythia}}\xspace}

\def\geant      {\mbox{\textsc{Geant4}}\xspace}

\def\photos     {\mbox{\textsc{Photos}}\xspace}

\def\tell1  {TELL1\xspace}
\def\ukl1   {UKL1\xspace}

\def\ptupsilon {\mbox{$p_{\rm T}^{\PUpsilon}$}\xspace}

\def\ptg {\mbox{$p_{\rm T}^{\g}$}\xspace}

\usepackage{cite} 
\usepackage{mciteplus}

\usepackage{longtable} 

\begin{document}

\renewcommand{\thefootnote}{\fnsymbol{footnote}}
\setcounter{footnote}{1}


\begin{titlepage}
\pagenumbering{roman}

\vspace*{-1.5cm}
\centerline{\large EUROPEAN ORGANIZATION FOR NUCLEAR RESEARCH (CERN)}
\vspace*{1.5cm}
\hspace*{-0.5cm}
\begin{tabular*}{\linewidth}{lc@{\extracolsep{\fill}}r}
\ifthenelse{\boolean{pdflatex}}
{\vspace*{-2.7cm}\mbox{\!\!\!\includegraphics[width=.14\textwidth]{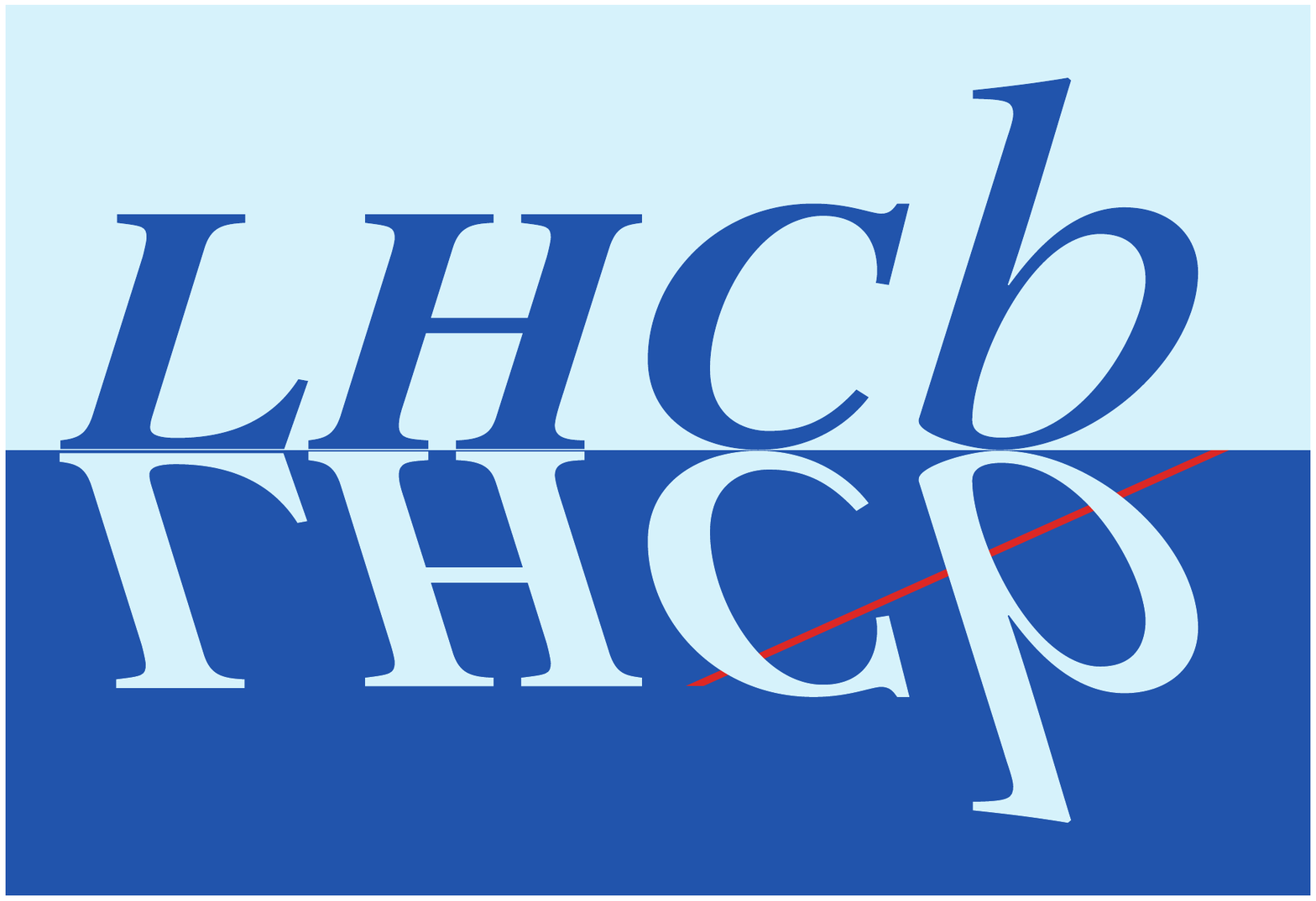}} & &}%
{\vspace*{-1.2cm}\mbox{\!\!\!\includegraphics[width=.12\textwidth]{lhcb-logo.eps}} & &}%
\\
 & & CERN-PH-EP-2014-206 \\  
 & & LHCb-PAPER-2014-040 \\  
 & & 4 September 2014 \\ 
 & & \\
\end{tabular*}

\vspace*{1.0cm}

{\bf\boldmath\huge
\begin{center}
   Measurement of the $\chib(3P)$ mass and of the relative rate of $\chibone(1P)$ and $\chibtwo(1P)$ production
\end{center}
}

\vspace*{1.5cm}

\begin{center}
The LHCb collaboration\footnote{Authors are listed at the end of this paper.}
\end{center}

\vspace{\fill}

\begin{abstract}
  \noindent
  The production of $\chi_b$ mesons in proton-proton collisions is studied using a data sample collected by the LHCb detector, 
at centre-of-mass energies of $\sqrt{s}=7$ and $8\tev$ and corresponding to an integrated luminosity of 3.0\invfb.
  The $\chi_b$ mesons are identified through their decays to $\OneS\g$ and $\TwoS\g$ using photons that converted to $e^+e^-$ pairs in the detector.
The relative prompt production rate of $\chibone(1P)$ and $\chibtwo(1P)$ mesons is measured
as a function of the \OneS transverse momentum in the $\chi_b$ rapidity range $2.0<y<4.5$. 
A precise measurement of the $\chib(3P)$ mass is also performed. Assuming a mass splitting between the $\chibone(3P)$ and the $\chibtwo(3P)$ states 
of 10.5~\mevcc, the measured mass of the $\chibone(3P)$ meson is
\begin{equation*}
m(\chibone(3P))= 10515.7^{+2.2}_{-3.9}\stat ^{+1.5}_{-2.1}\syst \mevcc. 
\end{equation*}
\end{abstract}

\vspace*{1.0cm}

\begin{center}
Submitted to JHEP
\end{center}

\vspace{\fill}

{\footnotesize 
\centerline{\copyright~CERN on behalf of the \lhcb collaboration, license \href{http://creativecommons.org/licenses/by/4.0/}{CC-BY-4.0}.}}
\vspace*{2mm}

\end{titlepage}


\newpage
\setcounter{page}{2}
\mbox{~}

\cleardoublepage


\renewcommand{\thefootnote}{\arabic{footnote}}
\setcounter{footnote}{0}


\pagestyle{plain} 
\setcounter{page}{1}
\pagenumbering{arabic}

%

\section{Introduction}
\label{sec:Introduction}
The study of production and properties of heavy quark-antiquark bound states (quarkonia) provides an important test of the underlying mechanisms described by quantum chromodynamics (QCD).
The quarkonium  (\ccbar and \bbbar) states in which quarks have parallel spins include the $S$-wave (\jpsi, $\PUpsilon$) and the $P$-wave (\chic, \chib) states, 
where each of the latter comprises a closely spaced triplet of $J=0,1,2$ spin states ($\chi_{cJ}$, $\chi_{bJ}$).
In high-energy proton-proton collisions at the \lhc, \qqbar pairs ($q=c,b$) are expected to be produced predominantly via a hard gluon-gluon interaction 
followed by the formation of bound quarkonium states. The production of the \qqbar pair is described by perturbative QCD,
while non-perturbative QCD is needed for the description of the evolution of the \qqbar pair to the bound state.
Several models have been developed for this non-perturbative part such as 
the colour singlet model~\cite{CSMBaier,CSMLikhoded,CSMBerger} and the non-relativistic QCD (NRQCD) model~\cite{NRQCD0,*NRQCD0err,NRQCD}, which also includes
the production of quarkonium via the colour octet mechanism.
Recent studies support the leading role of the colour singlet mechanism~\cite{Lansberg,Tramontano}. 
Measurements of the relative rate of $J=1$ and $J=2$ states provide information on the colour octet contribution.
This relative rate is also predicted to have the same dependence on the meson transverse momentum ($\pt$) in $\chi_b$ and $\chi_c$ states,
 once the $\pt$ of the $\chi_b$ meson is scaled by the ratio of $\chi_c$ and $\chi_b$ masses~\cite{Likhoded_chib}. 

Measurements of $\chi_c$ production and the ratio of the \chicone and \chictwo production cross-sections have been made previously 
using various particle beams and energies~\cite{WA11chic,HERABchic,CDFchic,CMSchic,LHCb-PAPER-2013-028}.
All the $\chi_b$ states are below the \BBbar threshold (where $B$ stands for $b$ mesons) and therefore can be studied 
through their radiative decays to the $\PUpsilon$ mesons, in the same way as the $\chi_c$ states were studied through their radiative decays to the \jpsi meson~\cite{LHCb-PAPER-2013-028}.

In this paper we report a measurement of the ratio of $\chibtwo(1P)$ to $\chibone(1P)$ production cross-sections 
$\sigma(pp\to\chibtwo(1P) X)/\sigma(pp\to\chibone(1P) X)$
at centre-of-mass energies of $\sqrt{s}=7$ and 8~\tev in the rapidity range $2.0<y<4.5$ as a function of the \OneS transverse momentum from 5 to 25\gevc.
The full LHCb sample is used, corresponding to an integrated luminosity of 3.0\,fb$^{-1}$.
The observation in LHCb data of the recently observed $\chib(3P)$ state~\cite{ATLASchib,D0chib}  is also presented.
The measurement of its mass and of the mass splitting between the $\chi_{bJ}(1P)$ states ($J=1$ and $J=2$)
provide useful information for testing QCD models~\cite{thchibmass,Rosner-Kwong,Ferretti-Galata}. 

The kinematically allowed transitions $\chib(1P)\to\OneS\g$, $\chib(2P)\to\OneS\g$, $\chib(3P)\to\OneS\g$ and $\chib(3P)\to\TwoS\g$ are studied. 
The $\PUpsilon(mS)$ ($m=1,2$) meson is reconstructed in the dimuon final state 
and only photons that convert in the detector material are used. 
The converted photons are reconstructed using \ep and \en tracks, allowing a 
separation of the $\chi_{b1}$ and $\chi_{b2}$ mass peaks, due to the improved 
energy resolution of converted photons with respect to that of photons identified with the calorimeter.
Any contribution from the $\chi_{b0}$ mesons decays is neglected, as their radiative decay rate 
is expected to be suppressed by an order of magnitude compared to that of the $\chi_{b2}$ meson~\cite{PDG2012,Rosner-Kwong}.

\section{Detector and data samples}
\label{sec:Detector}
The \lhcb detector~\cite{Alves:2008zz} is a single-arm forward
spectrometer covering the \mbox{pseudorapidity} range $2<\eta <5$,
designed for the study of particles containing \bquark or \cquark
quarks. The detector includes a high-precision tracking system
consisting of a silicon-strip vertex detector (VELO) surrounding the $pp$
interaction region, a large-area silicon-strip detector station located
upstream of a dipole magnet with a bending power of about
$4{\rm\,Tm}$, and three stations of silicon-strip detectors and straw
drift tubes placed downstream of the magnet.
The tracking system provides a measurement of momentum, \ptot,  with
a relative uncertainty that varies from 0.4\% at low momentum to 0.6\% at 100\gevc.
The total material before the first tracking station corresponds to about $25\%$ of a radiation length.   
The minimum distance of a track to a primary vertex, the impact parameter, is measured with a resolution of $(15+29/\pt)\mum$,
where \pt is in \gevc.
Different types of charged hadrons are distinguished using information
from two ring-imaging Cherenkov detectors. Photon, electron and
hadron candidates are identified by a calorimeter system consisting of
scintillating-pad and preshower detectors, an electromagnetic
calorimeter (ECAL) and a hadronic calorimeter. 
The reconstruction of converted photons is described in Sec.~\ref{sec:Selection}.
Muons are identified by a system composed of alternating layers of iron and multiwire
proportional chambers.

The LHCb coordinate system is right-handed with its origin at the nominal interaction point, the
$z$ axis aligned along the beam line towards the magnet and the
$y$ axis pointing upwards. The magnetic field is oriented along the $y$ axis.

The trigger consists of a
hardware stage, based on information from the calorimeter and muon
systems, followed by a software stage, which applies a full event
reconstruction.
Events used in this analysis are first required to pass a hardware trigger that selects muon candidates with  $\pt>1.76\gevc$ or dimuon candidates with a product of their \pt larger than
$(1.6~\gevcnosp)^2$.
In the  software trigger both muons are required to have
$\pt>0.5\gevc$, total momentum $\ptot>6\gevc$, and dimuon invariant mass greater than $4.7\gevcc$.

In the simulation, $pp$ collisions are generated using
\pythia~\cite{Sjostrand:2006za,*Sjostrand:2007gs} 
 with a specific \lhcb
configuration~\cite{LHCb-PROC-2010-056}.  Decays of hadronic particles
are described by \evtgen~\cite{Lange:2001uf}, in which final state
radiation is generated using \photos~\cite{Golonka:2005pn}. The
interaction of the generated particles with the detector and its
response are implemented using the \geant toolkit~\cite{Allison:2006ve, *Agostinelli:2002hh} as described in
Ref.~\cite{LHCb-PROC-2011-006}.
The simulated samples consist of events containing at least one $\PUpsilon$ meson 
that is forced to decay to two muons.
In a sample used for background studies, no restriction on the $\PUpsilon$ meson production mechanism is imposed. 
This sample is referred to as {\it inclusive $\PUpsilon$} in the following.
In another sample, used for the estimation of signal efficiencies and parametrisation, the 
$\PUpsilon$ is required to originate from a $\chib$ meson. This simulated sample is about 10 times larger than the data sample.

\section{Event reconstruction and selection}
\label{sec:Selection}
The reconstruction and selection of $\chi_b$ candidates closely follows 
Ref.~\cite{LHCb-PAPER-2013-028}.
Photons that convert in the detector material are reconstructed from pairs of oppositely charged electron candidates. 
Since the acceptance is lower for photons that convert in the VELO and the energy resolution is worse,
only $\gamma\to e^+e^-$ candidates without VELO hits are considered. This selection strongly favours
conversions that occur between the downstream end of the VELO and the first tracking station upstream of the magnet.
The $e^+e^-$ candidates are required to be within the ECAL acceptance and to produce electromagnetic clusters that have compatible 
coordinates in the non bending plane.
Any photon whose position in the ECAL is compatible with a straight line extrapolation of the electron track from the first tracking station is 
considered as a bremstrahlung photon. Its energy is added to the electron energy. 
If the same bremsstrahlung candidate is found for both the \ep and the \en, the photon energy is added randomly to one of the tracks.
The \ep and \en tracks (corrected for bremsstrahlung) are then extrapolated backwards in order to determine the conversion point and a vertex fit 
is performed  to reconstruct the photon momentum.
The transverse momentum of the photon candidate (\ptg) is required to be larger than 600\mevc and 
the invariant mass of the $\ep\en$ pair is required to be less than 50\mevcc, which removes most of the combinatorial background.
The resulting purity of the photon sample is determined from simulation to be about $99\%$.

\begin{figure}[tb]
  \begin{center}
    \includegraphics[width=0.7\linewidth]{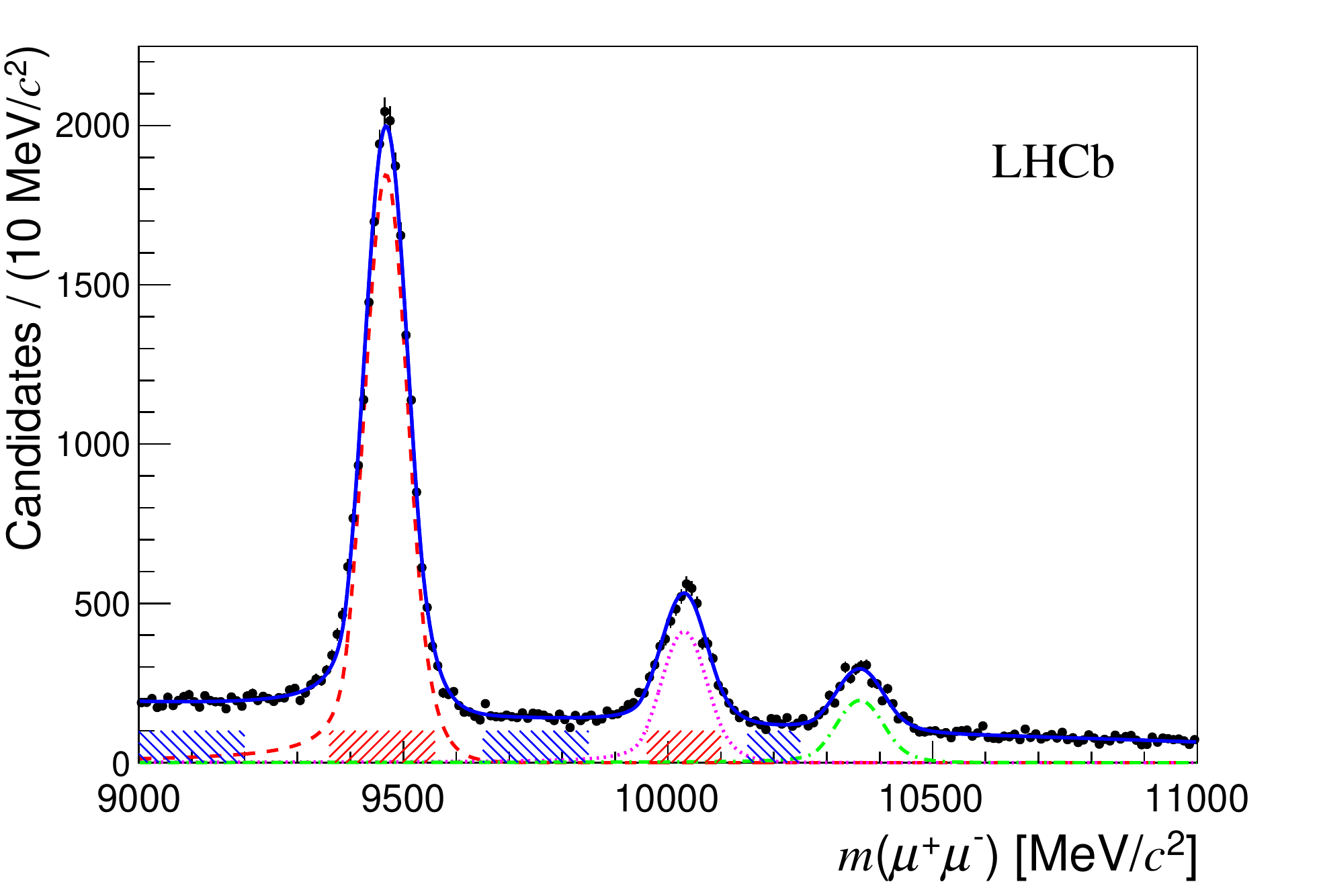}
    \vspace*{-.5cm}
  \end{center}
  \caption{
    \small Invariant dimuon mass of the $\PUpsilon$ candidates after the event selection requirements and before the $\PUpsilon$ mass range requirement. 
The distribution is fitted with the sum (blue line) of a double-sided Crystal Ball function for each 
$\PUpsilon$ state (dashed red line for $\OneS$, dotted pink line for $\TwoS$, 
dash-dotted green line for $\ThreeS$) and a second-order polynomial for the background (not shown). The hatched red  bands show the 
 signal regions and the hatched blue bands 
show the mass sidebands used for background studies.}
  \label{fig:mUpsilon}
\end{figure}

The $\PUpsilon$ candidate is reconstructed in its decay to the $\mumu$ final state. Each track must be identified as a muon with 
$\pt>2\gevc$ and $\ptot>8\gevc$. 
The two muons must originate from a common vertex with vertex fit $\chisq/\rm{ndf}$ smaller than 25. 
Only $\PUpsilon$ candidates with transverse momentum (\ptupsilon) greater than 4~\gevc are kept.
Figure~\ref{fig:mUpsilon} shows the invariant mass of $\PUpsilon$ candidates. The mass resolution is about 43~\mevcc.
The accepted mass ranges for the $\OneS$  and for the $\TwoS$ candidates are given in Table~\ref{tbl:trcuts}.

The $\PUpsilon$ and \g candidates are each associated with the primary vertex (PV) relative to which they have the smallest impact parameter 
$\chisq$, defined as the difference between the \chisq of the PV reconstructed with and without the considered tracks.
They are then combined to form a $\chi_b$ candidate. The $\chi_b$ decay time has to be smaller than 0.1~ps (about 5 times the observed resolution).
Loose requirements are applied in order to reject combinatorial background and poorly reconstructed candidates using the following variables:
the difference in $z$-positions of the primary vertices associated with the $\PUpsilon$ and \g candidates,
the \chisq of the $\chi_b$ candidate vertex fit  and the difference between the \chisq of the PV fitted with and without the $\chi_b$ candidate.
 These requirements remove about $30\%$ of the background and $8\%$ of the signal.
The cosine of the angle between the photon momentum  in the $\chi_b$ rest frame and the $\chi_b$ momentum is required to be positive.
This requirement halves the background while preserving $92\%$ of the signal.
The $\chi_b$ candidates are selected in the rapidity range $2.0<y<4.5$.

The $\chi_b$ candidates' mass is defined as $m^*(\mup\mun\g)\equiv m(\mup\mun\g)-m(\mup\mun)+m(\PUpsilon)$, where $m(\OneS)=9460.3\pm0.3$~\mevcc and 
$m(\TwoS)=10023.3\pm0.3$~\mevcc are the known $\PUpsilon$ mass values~\cite{PDG2012}.
This allows a nearly exact cancellation of the uncertainty due to the $\PUpsilon$ mass resolution and any possible bias on the $\PUpsilon$ 
candidates mass. 
The $\chi_b$ mass resolution is therefore dominated by the resolution on the photon energy.
The requirements on \ptupsilon and \ptg and the $\PUpsilon$ signal mass ranges used for each $\chib(nP)\to\PUpsilon(mS)\g$ decay mode 
are given in Table~\ref{tbl:trcuts}.
%
\begin{table}
\begin{center}
\caption{Selection criteria  for each $\chib(nP)\to\Upsilon(mS)\g$ transition. SB indicates sideband.\label{tbl:trcuts}}
{\renewcommand{\arraystretch}{1.2}
\begin{tabular}{l c c c c}
  $(n,m)$                 & (1,1) & (2,1) & (3,1) & (3,2)\\
\hline
\ptupsilon (\gevc) & $>4.0$  & $>4.0$  & $>5.0$ & $>6.0$ \\
\ptg (\gevc) & $>0.6$   & $>0.9$  & $>1.3$ & $>0.7$\\
\hline
$\PUpsilon$ mass range  (\mevcc) & \multicolumn{3}{ c}{$9360<m({\mup\mun})<9560$} & $9960<m({\mup\mun})<10100$\\
Low mass SB range  (\mevcc)      & \multicolumn{3}{ c}{$9000<m({\mup\mun})<9200$} & $9650<m({\mup\mun})<9850$\\
High mass SB range  (\mevcc)     & \multicolumn{3}{ c}{$9650<m({\mup\mun})<9850$} & $10150<m({\mup\mun})<10250$\\
\end{tabular}
}
\end{center}
\end{table}

\section{Sample composition and fit model}
\label{sec:fit}
Two background sources are considered in the sample of $\chi_b$ candidates.
One source is the non-$\PUpsilon$ background originating mainly from the Drell-Yan process where the dimuon pair is combined with a photon. 
The second source is the combinatorial background where a genuine $\PUpsilon$ is combined with a random photon.
The functions used for the fits are the sums of a background and signal functions.

The $\chi_{b1}$ and $\chi_{b2}$ peaks are each parametrised with a double sided Crystal Ball (CB) function~\cite{Skwarnicki:1986xj}: 
\begin{eqnarray}
\label{eq:CB}
\mathrm{CB}_i(m^*)\propto & {\rm exp}(-\frac{1}{2}(\frac{m^*-m_i}{\sigma_i})^2) & ~\mbox{ if }~ -\alpha_{\mathrm{L}}<\frac{m^*-m_i}{\sigma_i}<\alpha_\mathrm{R}  \nonumber
\\
\mathrm{CB}_i(m^*)\propto & \frac{(n_{\mathrm{L}}/\alpha_{\mathrm{L}})^{n_{\mathrm{L}}} {\rm exp}(-\frac{1}{2}\alpha_L^2)}{(n_{\mathrm{L}}/\alpha_{\mathrm{L}}-\alpha_{\mathrm{L}}-(m^*-m_i)/\sigma_i)^{n_{\mathrm{L}}}} & ~\mbox{ if }~ \frac{m^*-m_i}{\sigma_i}<-\alpha_{\mathrm{L}}
\\
\mathrm{CB}_i(m^*)\propto & \frac{(n_{\mathrm{R}}/\alpha_{\mathrm{R}})^{n_{\mathrm{R}}} {\rm exp}(-\frac{1}{2}\alpha_{\mathrm{R}}^2)}{(n_{\mathrm{R}}/\alpha_{\mathrm{R}}-\alpha_{\mathrm{R}}+(m^*-m_i)/\sigma_i)^{n_{\mathrm{R}}}} & ~\mbox{ if }~ \frac{m^*-m_i}{\sigma_i}>\alpha_{\mathrm{R}} \nonumber,
\end{eqnarray}
where the index $i=1(2)$ refers to the $\chi_{b1}$ ($\chi_{b2}$) CB function. 
The CB left tail accounts for events with 
unreconstructed bremsstrahlung, while the right tail accounts for events with overcorrected bremsstrahlung.
Simulation shows that the same tail parameters $\alpha_{\mathrm{R}}$ and $n_{\mathrm{L,R}}$ can be used for all the $\chi_{bi}(nP)$ states,
 $n_{\mathrm{L}}=n_{\mathrm{R}}=2.5$ and $\alpha_{\mathrm{R}}=1.0$, while  different values of $\alpha_{\mathrm{L}}$ have to be used: 
$\alpha_{\mathrm{L}}=0.20$, $0.25$ and $0.30$, 
for the $\chi_{bi}(1P)$, $\chi_{bi}(2P)$ and $\chi_{bi}(3P)$ shapes, respectively.
Since in the study of $\chi_c$ states it was found that 
the CB tail parameters were similar in data and simulation~\cite{LHCb-PAPER-2013-028},
 the values found with simulation are used for the $\chib$.
The CB width, $\sigma$, increases with the mass difference between the considered $\chi_b$  and $\PUpsilon$ states.
Fits to the mass distributions of $\chib(1P)\to\OneS\g$ and $\chib(2P)\to\OneS\g$ candidates indicate that the width 
is $10\%-20\%$ larger in data than in simulation. 
Therefore, the CB width is fixed to the value found with simulated events increased by $10\%$ and it is varied by $\pm10\%$ for
studies of the systematic effects. 

The shape of the non-$\PUpsilon$ background and its amplitude are estimated using the $\PUpsilon$ mass sidebands shown in Fig.~\ref{fig:mUpsilon} 
and given in Table~\ref{tbl:trcuts}.
The mass distribution of these candidates is fitted with an empirical function
\begin{equation}
\label{eq:bkgfunc}
f_{\rm bkg}(m^*)\propto{\arctan}\biggl(\frac{m^* -m_0}{c}\biggr)+b\biggl(\frac{m^*}{m_0}-1\biggr)+a \; ,
\end{equation} 
where $m_0$, $a$, $b$ and $c$ are free parameters.
This function is then used to parametrise the non-$\PUpsilon$ background contribution with all parameters fixed to the fitted values.
The shape of the combinatorial background is estimated using the inclusive $\PUpsilon$ simulated sample and parametrised with Eq.~(\ref{eq:bkgfunc}).
All parameters are fixed to the values found with simulation except for the normalisation. In the case of the $\chib(3P)\to\TwoS\g$ 
transition, this shape 
does not reproduce the data properly and the value of the $m_0$ parameter is therefore left free in the fit. 
This discrepancy is due to 
mismodeling of the \ptupsilon spectrum in simulation
and is accounted for in the systematic uncertainties.

The fits have at most six free  parameters: the mean mass value for the $\chibone$ peak $m_1$,
 the mass difference between the $\chibtwo$ and $\chibone$ peaks $\Delta m_{12}$,
the normalisation of the $\chibone$ CB function $A_1$,
 the ratio of the $\chibtwo$ to $\chibone$ CB amplitudes $r_{12}$, the normalisation of the combinatorial background $A_{comb}$ and
the $m_0$ parameter for the combinatorial background shape.

\section{$\boldsymbol{\chib}$ meson masses}
\label{sec:mass}
\subsection{Mass measurements}
\begin{figure}[htpb]
  \begin{center}
\begin{tabular}{cc}
    \hspace{-.8cm}\includegraphics[width=0.58\linewidth]{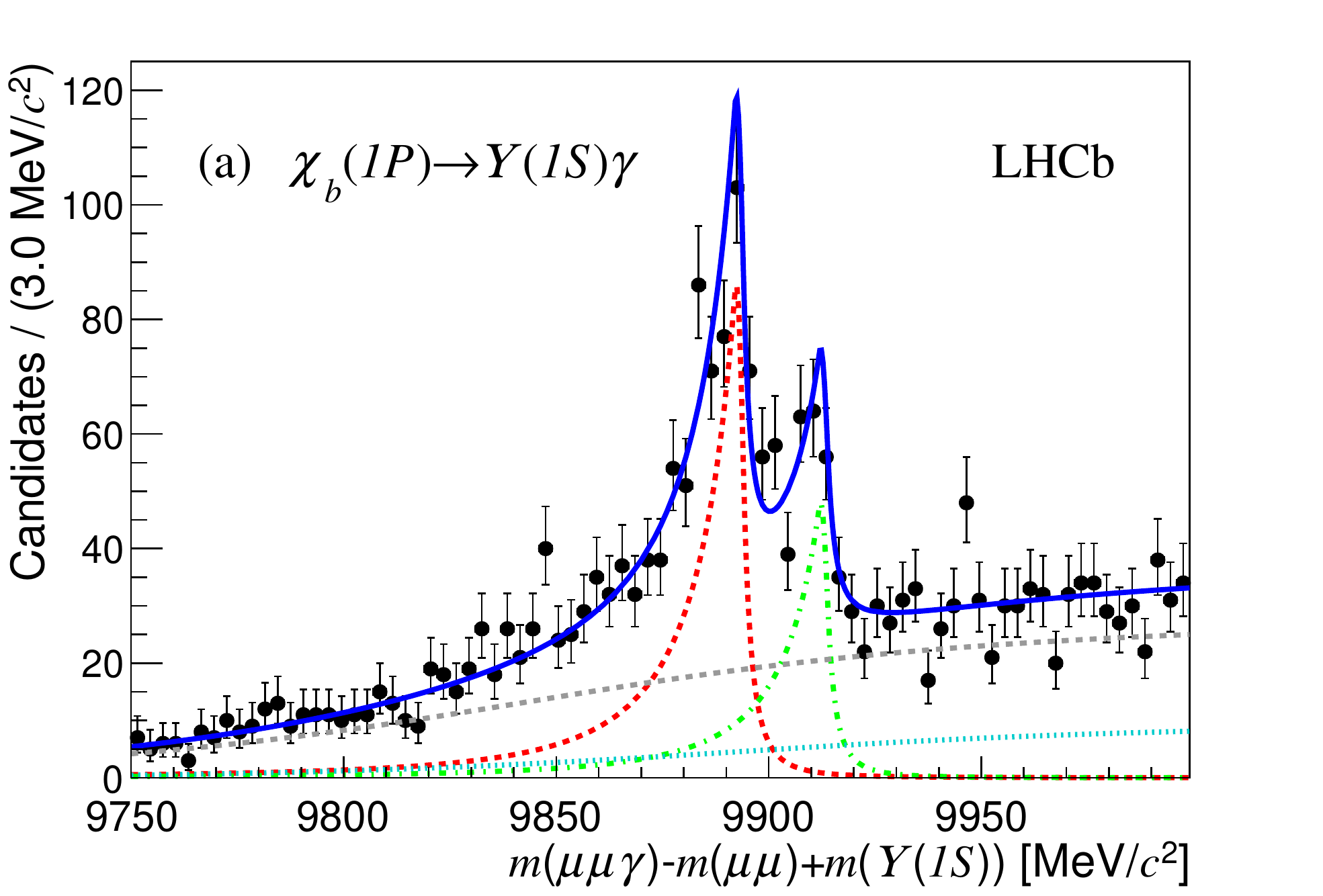} &
    \hspace{-1.cm}\includegraphics[width=0.58\linewidth]{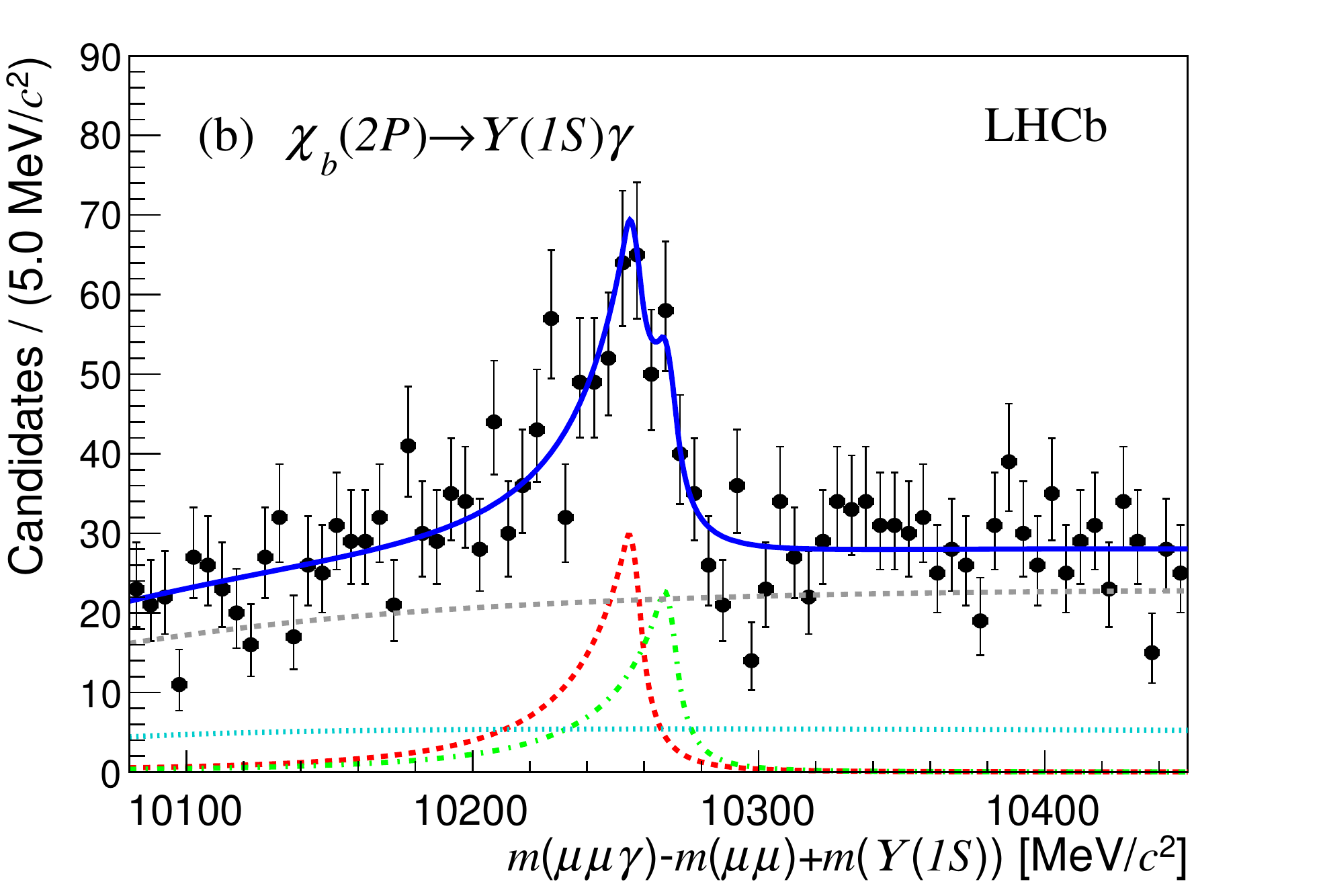} \\
    \hspace{-.8cm}\includegraphics[width=0.58\linewidth]{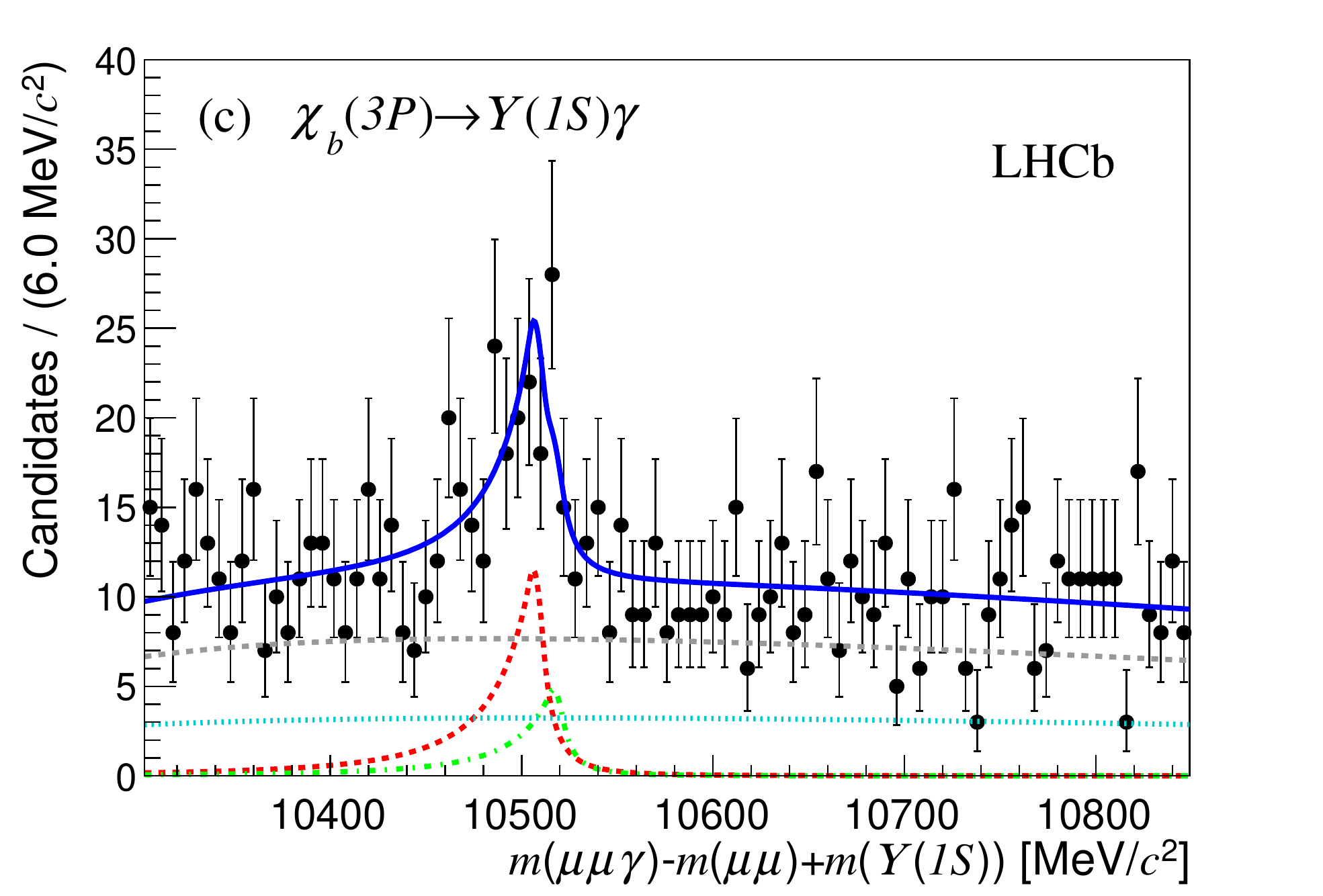} &
    \hspace{-1.cm}\includegraphics[width=0.58\linewidth]{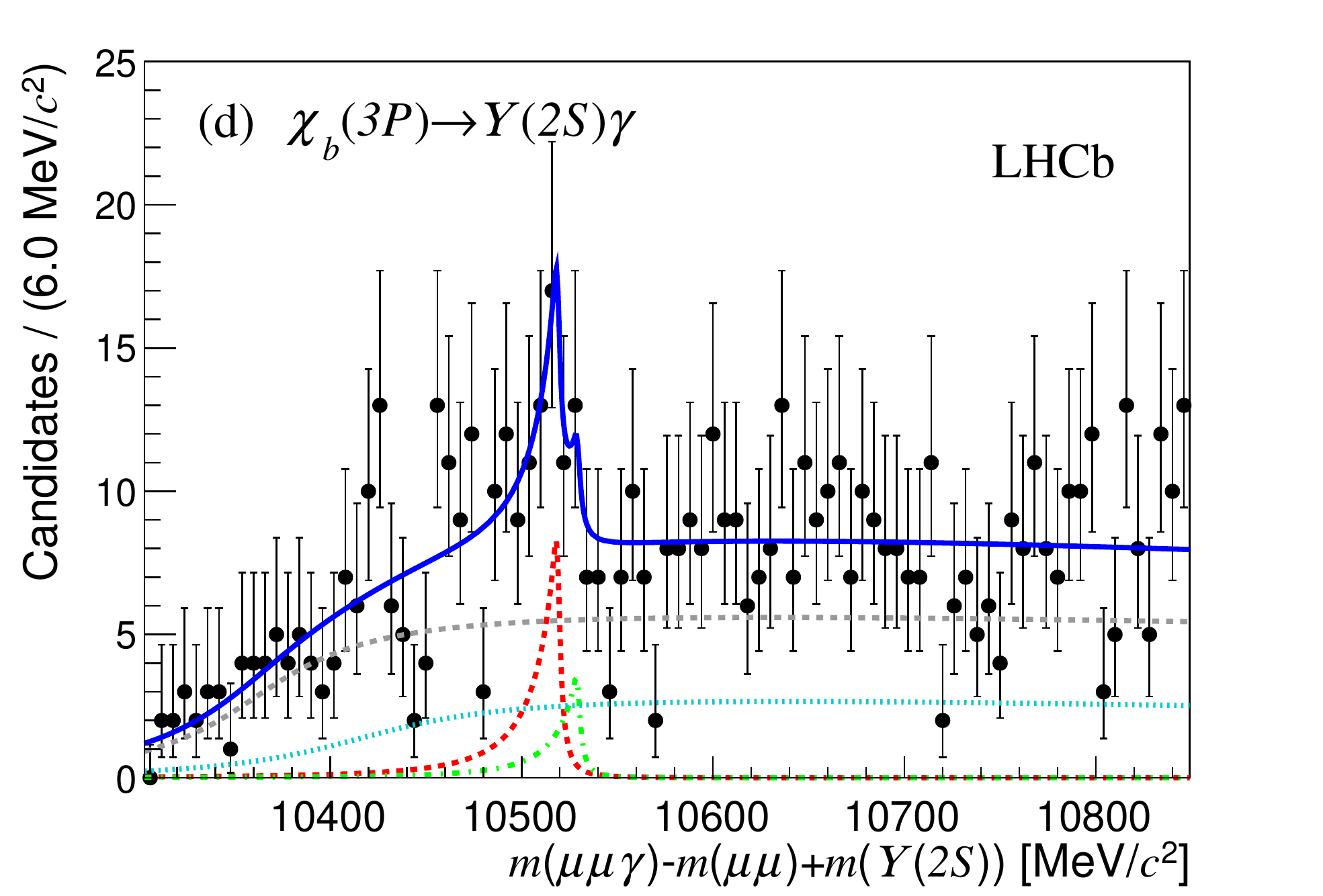} \\
    \hspace{-.8cm}\includegraphics[width=0.58\linewidth]{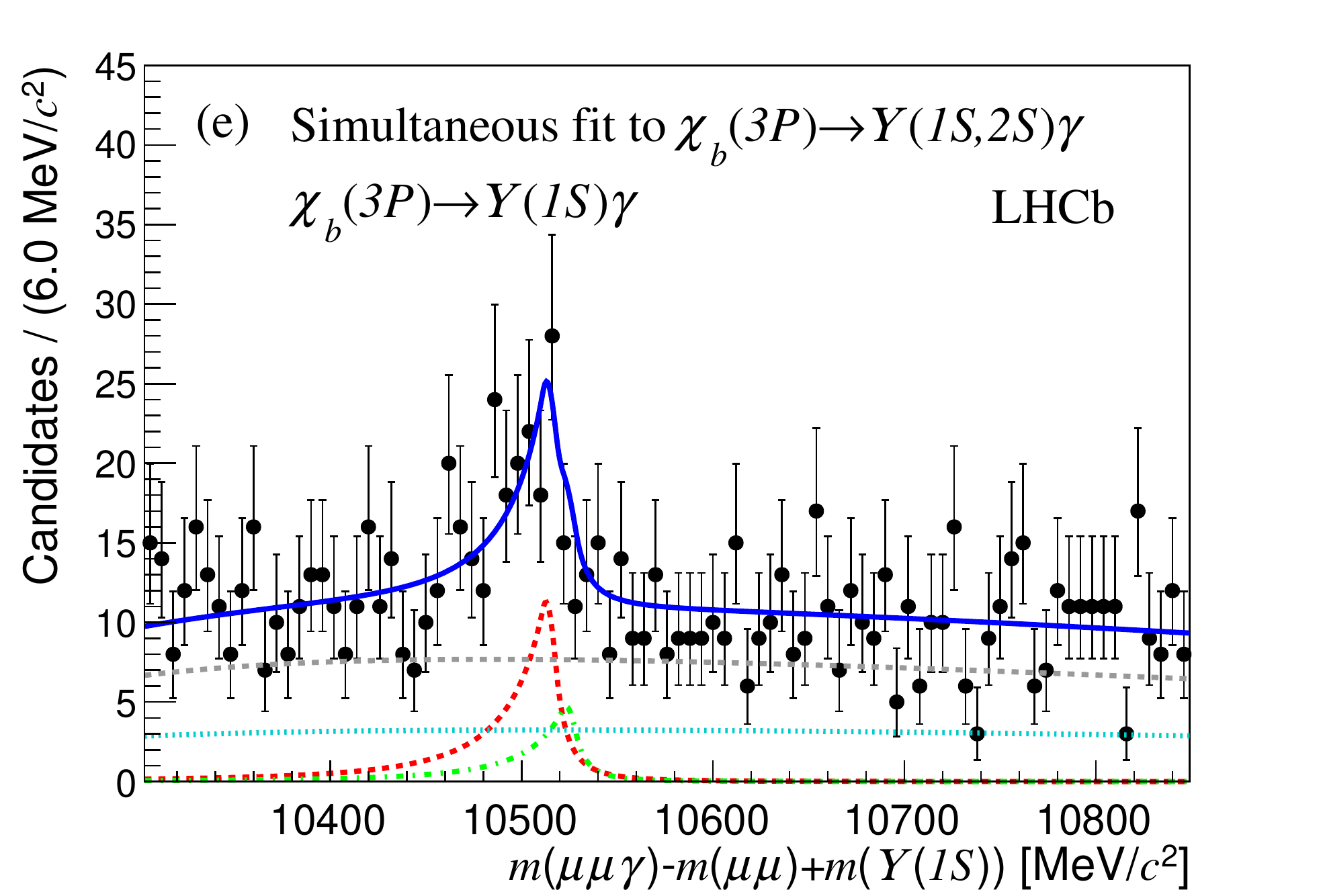} &
    \hspace{-1.cm}\includegraphics[width=0.58\linewidth]{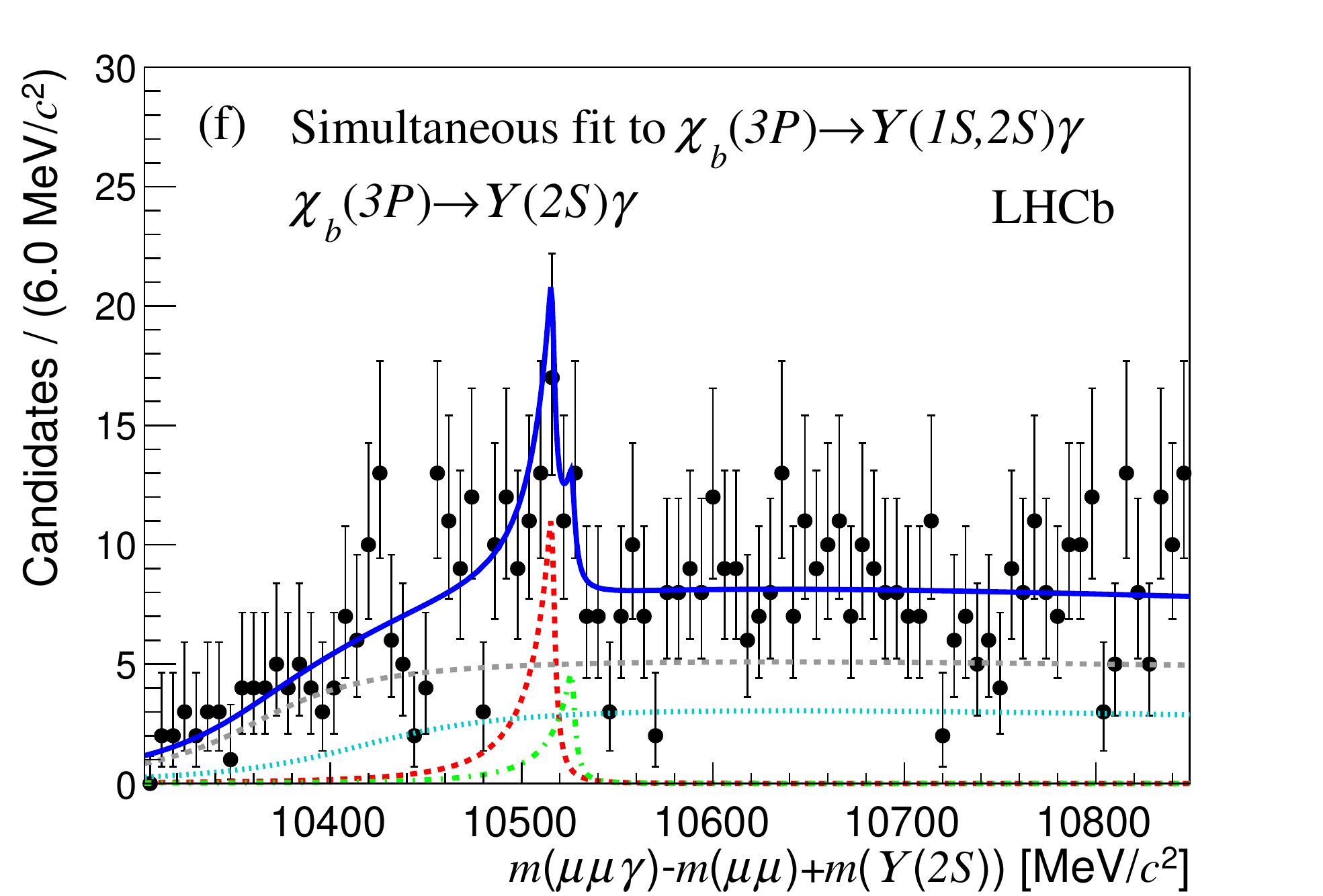} \\
\end{tabular}
    \vspace*{-.5cm}
  \end{center}
  \caption{
    \small Distribution of $m^*(\mup\mun\g)\equiv m(\mup\mun\g)-m(\mup\mun)+m(\PUpsilon)$ for $\chib$ candidates with fit projections overlaid 
for (a) $\chib(1P)\to\OneS\g$, (b) $\chib(2P)\to\OneS\g$, 
(c,e) $\chib(3P)\to\OneS\g$  and (d,f) $\chib(3P)\to\TwoS\g$  channels. The result of the simultaneous fit to the
  $\chib(3P)\to\OneS\g$ and $\chib(3P)\to\TwoS\g$ mass distributions is shown in (e) and (f).
The cyan dotted line shows the non-$\PUpsilon$ background, the grey dashed line shows the combinatorial background,
the red dashed line the $\chi_{b1}$ contribution, the green dash-dotted line the $\chibtwo$ contribution,
  and the blue full line the sum of all these contributions.
  \label{fig:Fit1P-3P}}
\end{figure}

The masses of the $\chib$ mesons are determined using unbinned maximum likelihood fits to the $\chib$ mass distributions using the parametrisation
 described in Sec.~\ref{sec:fit}.
Figures~\ref{fig:Fit1P-3P} (a) and (b) show the mass distributions for the $\chib(1P)\to\OneS\g$ and  $\chib(2P)\to\OneS\g$ decays with 
the fit results overlaid.
In these fits the free parameters are $m_1$, $A_1$, $\Delta m_{12}$, $r_{12}$ and $A_{comb}$.
Table~\ref{tbl:chibmass} reports the resulting mass determinations for these states compared to the world average values~\cite{PDG2012}.
A small bias is expected on the measured masses, attributed to unreconstructed bremsstrahlung of the $e^+e^-$ pair.
This bias is proportional to the Q-value of the transition and is expected, from simulation, to be about $-0.5$ and $-1.5$~\mevcc  for the 
$\chib(1P)\to\OneS\g$ and  $\chib(2P)\to\OneS\g$ decays, respectively. The measurements given in Table~\ref{tbl:chibmass} 
are not corrected for this bias and are consistent with such a bias.
On the other hand the $\chib(3P)$ mass measured using the $\chib(3P)\to\PUpsilon(mS)\g$ transitions is corrected for the bias estimated with simulation, 
$-3.0\pm2.0\mevcc$ and $-0.5\pm0.5$~\mevcc for $m=1$ and $m=2$, respectively, where the uncertainties
cover possible discrepancies between data and simulation.

In the case of the $\chib(3P)$ meson, the mass splitting and the relative yields are also fixed,
as the spin-1 and spin-2 peaks cannot be separated.
 Theory predictions vary from 9 to 12~\mevcc~\cite{thchibmass,Rosner-Kwong}
 for $\Delta m_{12}$ and this parameter is fixed to 10.5~\mevcc. 
The value of $r_{12}$ is fixed based on theoretical predictions~\cite{Rosner-Kwong} and our experimental measurement. 
It can be expressed as the product of the ratio of branching fractions to $\PUpsilon\g$ 
and of the ratio of production cross-sections of the $\chibtwo(3P)$ and $\chibone(3P)$ states.
Predictions for branching fractions are found in Refs.~\cite{Rosner-Kwong,Ferretti-Galata}.
The predictions from Ref.~\cite{Rosner-Kwong} agree well with the experimental measurements for the $\chib(1P)$ and the $\chib(2P)$ mesons.
The model of Ref.~\cite{Rosner-Kwong} predicts similar values for the two transitions,
 ${\cal B}(\chibtwo(3P)\to\PUpsilon(mS) \gamma))/{\cal B}(\chibone(3P)\to\PUpsilon(mS) \gamma)\approx 0.47$ ($m=1,2$).
According to Ref.~\cite{Likhoded_chib}  the ratio of production cross-sections is expected to be the same for the $\chib(3P)$ and $\chib(1P)$ mesons
and thus, using the measurement detailed in Sec.~\ref{sec:rates}, we obtain $\sigma(\chibtwo(nP))/\sigma(\chibone(nP))=0.9\pm0.2$.

To summarise, the value $r_{12}=0.47\times0.9=0.42$ is used in the fits to the mass distributions associated with 
the transitions of the $\chib(3P)$ meson to $\OneS$ and $\TwoS$ mesons.
Table~\ref{tbl:3Pmass} gives the result of the fits to the mass distributions for the $\chib(3P)\to\OneS\g$ and $\chib(3P)\to\TwoS\g$  transitions.
A simultaneous fit to these two distributions is also performed and the result is reported in the last column of Table~\ref{tbl:3Pmass}. 
Figure~\ref{fig:Fit1P-3P} shows the results of these fits.
The $\chib(3P)\to\OneS\g$ and $\chib(3P)\to\TwoS\g$ decays are seen with a statistical significance,
determined from the likelihood ratio of the fits with background only and with signal plus background
hypotheses, of $6.0\sigma$ and $3.6\sigma$ respectively. The total statistical significance determined with the simultaneous fit is $6.9\sigma$.
\begin{table}
\begin{center}
\caption{Fitted values of the $\chib(nP)$ $(n=1,2)$ masses (in \mevcc) from the $\chib(nP)\to\OneS\g$ transitions, 
compared to the world average values. The uncertainties are statistical only.
}
\label{tbl:chibmass}
{\renewcommand{\arraystretch}{1.2}
\begin{tabular}{l  c c}
$(n,m)$       & (1,1)             & (2,1)          \\
\hline
$m_1$       &  $9892.3\pm0.5$ & $10254.7\pm1.3$ \\
$m_1$ world average  &  $9892.8\pm0.4$ & $10255.5\pm0.6$ \\
\hline
$\Delta m_{12}$    &  $19.81\pm0.65$   & $12.3\pm2.6$     \\
$\Delta m_{12}$ world average & $19.43\pm0.37$     & $13.5\pm0.6$    \\
\end{tabular}
}
\end{center}
\end{table}
\begin{table}
\begin{center}
\caption{Fitted values of the $\chib(3P)$ mass (in \mevcc) for the $\chib(3P)\to\PUpsilon(mS)\g$ ($m=1,2$) transitions. 
The last column gives the result of the simultaneous fit to the two transitions. 
The values are corrected for the mass bias ($-3$~\mevcc and $-0.5$~\mevcc for the $\OneS$ and $\TwoS$ 
transitions, respectively). The last row gives the total $\chib$ yields. The uncertainties are statistical only.}
\label{tbl:3Pmass}
{\renewcommand{\arraystretch}{1.3}
\begin{tabular}{l  c c c}
%
$(n,m)$       &  (3,1)           & (3,2)     & (3,1)+(3,2)\\
\hline
$m_1$       &   $10509.0^{+5.0}_{-2.6}$ & $10518.5^{+1.9}_{-1.3}$ & $10515.7^{+2.2}_{-3.9}$\\
$\Delta m_{12}$ & $10.5$ (fixed) & $10.5$ (fixed) & $10.5$ (fixed) \\
$N(\chib)$  &   $107\pm19$              & $41\pm12$               &  $169\pm25$\\
\end{tabular}
}
\end{center}
\end{table}
\subsection{Systematic uncertainties}
\label{sec:systmass}
The systematic uncertainties on the measurement of the $\chib(nP)$ ($n=1,2$) mass splitting and of the $\chib(3P)$ mass 
are detailed as follows.

First the systematic uncertainties related to the signal parametrisation are considered.
The $\chi_{b0}$ contribution is expected to be small because its branching fraction to $\OneS\g$ is less than
$2\%$ for $\chib(1P)$ and $\chib(2P)$ mesons~\cite{PDG2012}.
In order to estimate the systematic uncertainty due to the presence
of a $\chi_{b0}$ or another unknown state, a third CB function is added to the fit, with a peak position fixed to the 
world average value for  the $\chib(nP)$ for $n=1,2$ and left free for the $\chib(3P)$. 
The resulting yield of $\chibzero$ mesons is  compatible with zero.
The Gaussian width of the CB function is varied within $\pm10\%$  to cover possible differences between data and simulation.
 For these two fit variations, the differences between results of the nominal and alternative fits are 
taken as systematic uncertainties, added in quadrature and referred to as {\it signal} uncertainty in Table~\ref{tbl:systmass}.

Imperfect modelling of the background is also considered as a possible source of systematic uncertainty.
The normalisation of the non-$\PUpsilon$ background is varied within the uncertainty of the estimated number of
background events under the $\PUpsilon$ peak (typically 10$\%$). 
Negligible variations are observed when the shape of this background is determined using only the low or the high mass sideband. 
Therefore no systematic uncertainty is assigned from the non-$\PUpsilon$ background modelling.
The shape of the combinatorial background is particularly sensitive to the $m_0$ value, therefore this parameter is varied within twice
its uncertainty. In the case of the $\chib(3P)\to\TwoS\g$ transition, where the value of $m_0$ is left free in the fit, 
the value found in simulation is used in an alternative fit, leading to a change of 0.1~\mevcc on the $\chib(3P)$ mass.
The fit range is also varied by $\pm100\mevcc$ on both sides. 
The differences between results of the nominal fit and these two alternative fits are
taken as systematic uncertainties and added in quadrature.
The resulting systematic uncertainty is referred to as {\it background} uncertainty.

The uncertainty on the mass bias (2.0 and 0.5~\mevcc for the $\chib(3P)$ mass measurement based on the transition to $\OneS$ and $\TwoS$ 
respectively) is assigned as systematic uncertainty.
For the simultaneous fit to the two $\chib(3P)$ mass distributions, the two biases are varied independently within their uncertainties
and the largest variation is taken as systematic uncertainty.
A small bias is expected on the $\chib(1P)$ mass splitting and is estimated to be at most 0.10~\mevcc, which is added
as a systematic uncertainty.
No significant bias on the $\chib(nP)$ mass splitting is expected from the fit procedure.

For the determination of the $\chib(3P)$ mass, the $\Delta m_{12}$ and $r_{12}$ parameters are fixed in the nominal fit.
They are varied independently within their expected uncertainties in order to evaluate the associated systematic uncertainties.
The mass splitting, $\Delta m_{12}$, is varied between 9 and 12 \mevcc  and the $r_{12}$ parameter is varied by $\pm 30\%$,
which includes theoretical uncertainties 
and the precision on the $\chib(1P)$ production ratio measured in this work and used to estimate $r_{12}$. 

Finally, the $0.3\mevcc$ uncertainty on the world-average values of the \OneS and  \TwoS masses is added as a systematic uncertainty to the  $\chib(3P)$ mass. 

Table~\ref{tbl:systmass} lists the individual systematic uncertainties. 
The total systematic uncertainty is the quadratic sum of all individual uncertainties.
%
\begin{table}
\begin{center}
\caption{Summary of the systematic uncertainties on the $\chib(nP)$ ($n=1,2$) mass splitting and on the $\chibone(3P)$ mass in \mevcc.
The last column refers to the simultaneous fit to the two transitions.}
\label{tbl:systmass}
{\renewcommand{\arraystretch}{1.2}
\begin{tabular}{l c c c c c}
 & $\Delta m_{12}(\rm{1P})$& $\Delta m_{12}(\rm{2P})$ & $m(\chibone(\rm{3P}))$ & $m(\chibone(\rm{3P}))$ &$m(\chibone(\rm{3P}))$\\
 &                         &                          & from $\OneS$   & from $\TwoS$        & combined\\
\hline
Signal  & $\pm0.16$ & $\pm0.5$ & $\pm0.3$ & $\pm0.1$ & $\pm0.6$\\
Background & $\pm0.08$ & $\pm0.3$ & $\pm0.2$ & $\pm0.1$ & $\pm0.2$\\
Bias  & $\pm0.10$   & $\pm0.1$ & $\pm2.0$ & $\pm0.5$ & $^{+1.2}_{-1.6}$\\
$r_{12}$ & -          &      - & $^{+0.7}_{-0.4}$ & $^{+0.1}_{-0.2}$& $^{+0.6}_{-1.1}$\\
$\Delta m_{12}$ & -   &      -     &$\pm1.2$ & $\pm0.1$ & $\pm0.3$\\
$m(\PUpsilon)$       &  -      &     -  & $\pm0.3$ & $\pm0.3$ &$\pm0.3$\\
\hline
Total & $\pm0.20$    & $\pm0.6$   & $^{+2.5}_{-2.4}$ & $\pm0.6$& $^{+1.5}_{-2.1}$\\
\end{tabular}
}
\end{center}
\end{table}

%
\section{Relative rate of $\boldsymbol{\chibtwo(1P)}$ and $\boldsymbol{\chibone(1P)}$ production \label{sec:rates}}
\subsection{Measurement of the relative rates}
\label{sec:rates1}
The production cross-section ratio of the $\chibtwo(1P)$ and $\chibone(1P)$ mesons is measured in 
three \ptupsilon ranges of different size (the bin limits are given in Table~\ref{tbl:rates}) using
\begin{equation}
\frac{\sigma(\chibtwo)}{\sigma(\chibone)}=\frac{N_{\chibtwo}}{N_{\chibone}}\frac{\varepsilon_{\chibone}}{\varepsilon_{\chibtwo}}\frac{\BR(\chibone\to\OneS\g)}{\BR(\chibtwo\to\OneS\g)},
\label{eq:rxs}
\end{equation}
where $\sigma(\chi_{bJ})$ ($J=1,2$) is the  $\chi_{bJ}(1P)$ meson production cross-section; $N_{\chi_{bJ}}$ is the  $\chi_{bJ}(1P)$ yield;
 $\varepsilon_{\chi_{bJ}}$ is the efficiency to trigger, detect, reconstruct and select a $\chi_{bJ}$ meson including the contribution from the
approximately  $20\%$ probability for a photon to convert 
upstream of the first tracking station; and
$\BR(\chibone(1P)\to\OneS\g)=(33.9\pm2.2)\%$ and 
$\BR(\chibtwo(1P)\to\OneS\g)=(19.1\pm1.2)\%$
are the known branching fractions~\cite{PDG2012} .

The inefficiency is dominated by the converted photon  acceptance and reconstruction: low-energy photons produce low-energy electrons, which
have a high chance to escape the detector due to the magnetic field. 
The efficiency of converted photon reconstruction and selection relative to non-converted photons is measured
in Ref.~\cite{LHCb-PAPER-2013-028} and ranges from about $1\%$ at \ptg of $600~\mevc$ to $3\%$ at \ptg of $2000~\mevc$. 
These numbers include the conversion probability.
Due to the correlation between the \pt of the photon and that 
of the $\PUpsilon$ meson, the efficiency is lower for low \ptupsilon.
The ratio of efficiencies is given in Table~\ref{tbl:rates}. This ratio differs from unity because the \ptupsilon spectrum is
different for $\chibone$ and $\chibtwo$ in \pythia8, as expected~\cite{Likhoded_chib}. 
The ratio of efficiencies is also calculated assuming equal \pt spectra. It is still slightly
different from unity due to the small difference in the $\chibone$ and $\chibtwo$ masses.

\begin{figure}[tb]
  \begin{center}
\begin{tabular}{cc}
    \hspace{-.8cm}\includegraphics[width=0.58\linewidth]{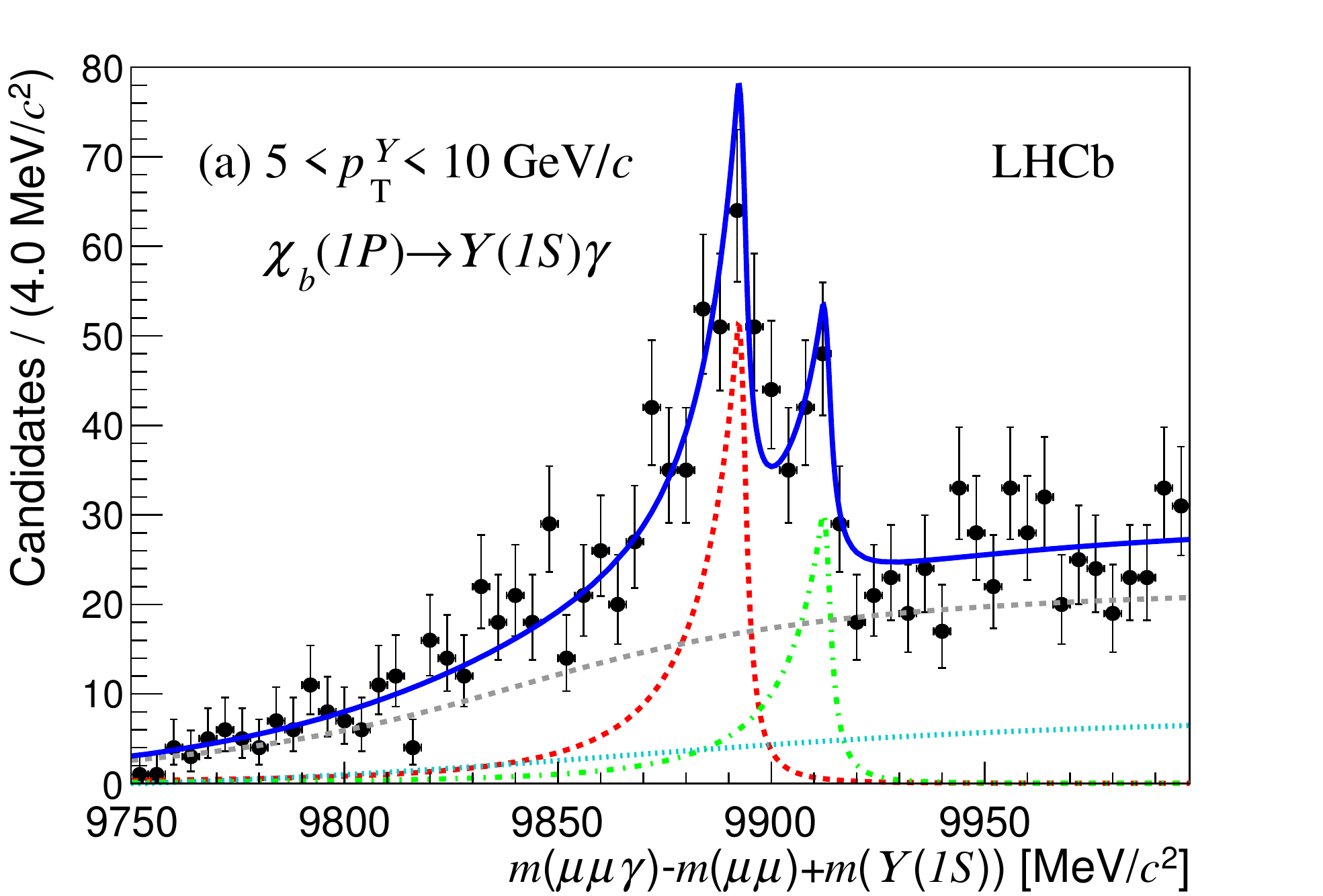} &
    \hspace{-1.cm}\includegraphics[width=0.58\linewidth]{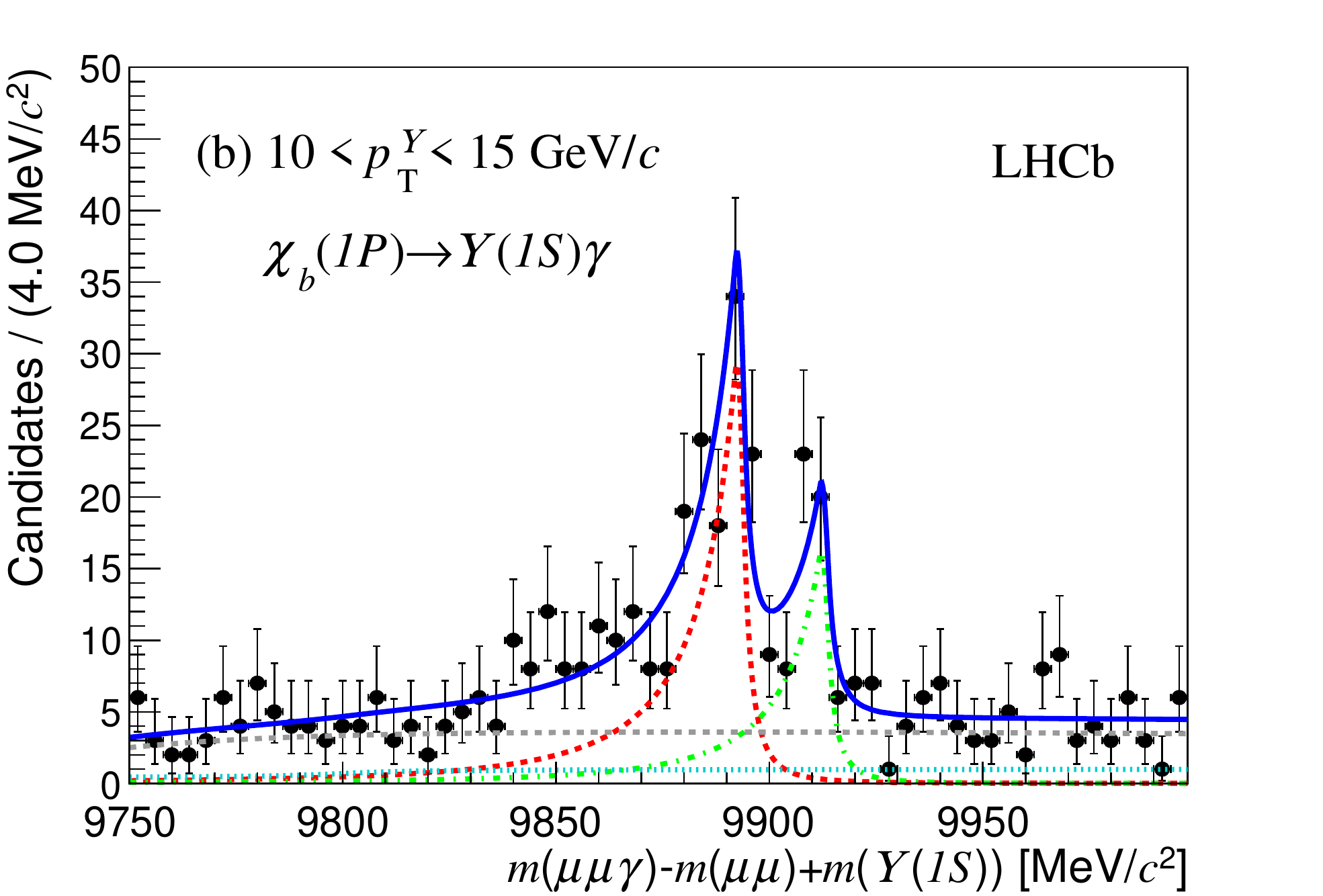} \\
    \hspace{-.8cm}\includegraphics[width=0.58\linewidth]{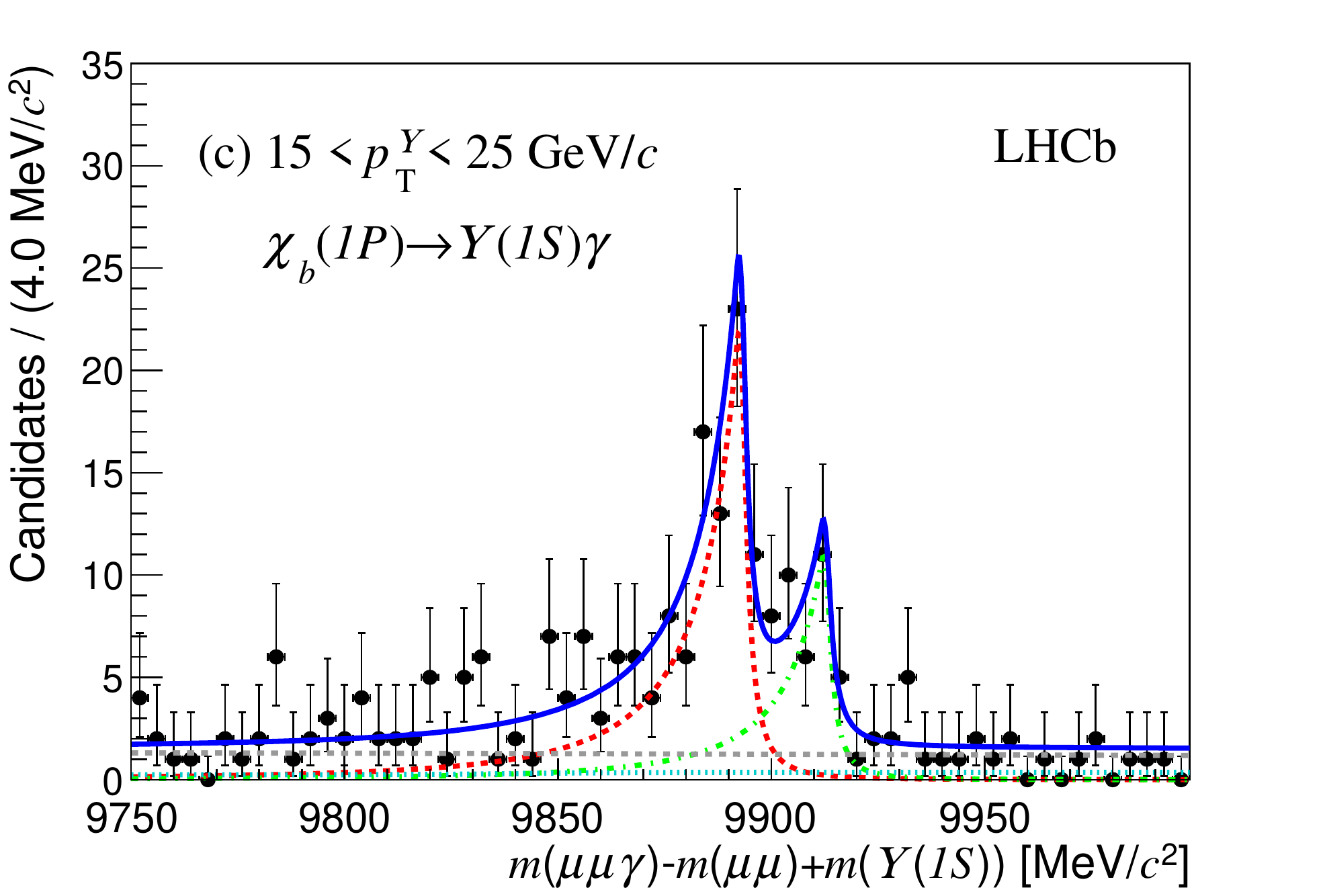} &\\
    \vspace*{-1.0cm}
\end{tabular}
  \end{center}
  \caption{
    \small Distribution of $m^*(\mup\mun\g)\equiv m(\mup\mun\g)-m(\mup\mun)+m(\PUpsilon)$ for $\chib(1P)$ candidates with fit projections overlaid 
for each of the three ranges in \ptupsilon: (a) 5--10 ~\gevc,  (b) 10--15~\gevc  and (c) 15--25~\gevc.
The cyan dotted line show the non-$\PUpsilon$ background, the grey dashed line shows the combinatorial background,
the red dashed line the $\chi_{b1}$ contribution, the green dash-dotted line the $\chi_{b2}$ contribution and the 
blue full line the sum of all these contributions.
  \label{fig:Fitrates}}
\end{figure}

The  mass distribution of $\chib$ candidates in each \ptupsilon bin is fitted using the signal and background 
functions described in Sec.~\ref{sec:fit}.
In these fits the mass of the $\chibone$ state and the mass splitting are fixed to the values found from the fit to the whole data set 
(see Table~\ref{tbl:chibmass}) and then varied within their uncertainties for systematic studies.
The result of the fit is shown in Fig.~\ref{fig:Fitrates} and the ratio of yields is given in Table~\ref{tbl:rates} for each \ptupsilon range.
\begin{table}
\begin{center}
\caption{Relative rate of $\chibone(1P)$ and $\chibtwo(1P)$ production and ratio of total efficiency (in the three \ptupsilon ranges). 
Uncertainties only refer to the statistical contributions.}
\label{tbl:rates}
{\renewcommand{\arraystretch}{1.1}
\begin{tabular}{l  c c c}
\ptupsilon bin (\gevc)& 5--10      &   10--15   & 15--25 \\
\hline
$N(\chibtwo)/N(\chibone)$ & $0.61\pm0.15$ & $0.57\pm0.15$ &$0.52\pm0.15$ \\
$\varepsilon(\chibone)/\varepsilon(\chibtwo)$ & $1.01\pm0.03$ & $0.90\pm0.05$ &$1.18\pm0.11$ \\
\end{tabular}
}
\end{center}
\end{table}
\subsection{Systematic uncertainties}
The same sources of systematic uncertainties as for the mass measurements (see Sec.~\ref{sec:systmass}) are investigated and reported in Table~\ref{tbl:systrates}.
Additional systematic checks relevant only for the relative rates of $\chibtwo(1P)$ and $\chibone(1P)$ are detailed as follows.

The dominant uncertainty on the ratio of efficiencies is due to the limited knowledge of the efficiency for reconstructing converted photons, which
is estimated following Ref.~\cite{LHCb-PAPER-2013-028} and amounts to $4\%$ on the relative rates. This uncertainty is added in quadrature to the 
uncertainty due to the limited size of the simulated sample.

Due to the large size of the \pt bins, the efficiency depends on the choice of the  \pt spectrum of $\chi_b$ production as discussed in Sec.~\ref{sec:rates1}.
In order to assess the uncertainty due to the shape of the \pt spectrum, the  simulated $\chi_{b2}$ ($\chi_{b1}$) spectrum 
is changed to be identical
to the simulated $\chi_{b1}$ ($\chi_{b2}$) spectrum. The relative difference in the ratio of efficiencies is taken as a 
systematic uncertainty.

The fit is also performed on simulated data and a mean bias of $(-4\pm4)\%$ is observed on the relative yields. 
A systematic uncertainty of $\pm 4\%$  is added to take the possible bias into account.
The values of the $\chibone(1P)$ mass $m_1$ and of the mass splitting $\Delta m_{12}$ are also varied within their uncertainties 
from Table~\ref{tbl:chibmass}. 
The variation of the result is taken as systematic uncertainty and is added in quadrature to the uncertainty referred to as {\it signal}. 

Table~\ref{tbl:systrates} lists the systematic uncertainties on the relative rates. 
The total systematic uncertainty is the quadratic sum of all individual uncertainties.
The ratio of cross-sections is also affected by the uncertainties on the branching fraction of $\chib(1P)\to\OneS\g$, leading to
an additional systematic uncertainty of $9.0\%$~\cite{PDG2012}.
\begin{table}
\begin{center}
\caption{Summary of the systematic uncertainties on the $\chib(1P)$ relative rates, expressed as fractions of the relative rate.}
\label{tbl:systrates}
{\renewcommand{\arraystretch}{1.2}
\begin{tabular}{l c c c}
\ptupsilon bin (\gevc)& 5--10      &   10--15   & 15--25 \\
\hline
Signal           & $\pm0.05$  & $\pm0.08$ & $\pm0.08$\\
Background         & $\pm0.06$ & $\pm0.04$ & $\pm0.03$  \\
Fit bias          & $\pm0.04$ & $\pm0.04$ & $\pm0.04$\\
Efficiency        & $\pm0.05$ & $\pm0.06$ & $\pm0.10$ \\
\pt model         & $ -0.13$  & $ -0.05$  & $ -0.04$ \\
\hline
Total             & $^{+0.10}_{-0.16}$& $^{+0.12}_{-0.13}$ & $^{+0.13}_{-0.14}$\\
\end{tabular}
}
\end{center}
\end{table}

\section{Results}
\label{results}
The results for the $\chib(1,2P)$ mass splittings between the $J=1$ and $J=2$ states 
\begin{equation*}
\Delta m_{12}(1P)= 19.81\pm0.65\stat\pm0.20\syst \mevcc
\end{equation*}
\begin{equation*}
\Delta m_{12}(2P)= 12.3\pm2.6\stat\pm0.6\syst \mevcc
\end{equation*}
are in agreement with the world average values, $\Delta m_{12}(1P)= 19.43\pm0.37 \mevcc$ and $\Delta m_{12}(2P)= 13.5\pm0.6 \mevcc$~\cite{PDG2012}.
A measurement of the $\chibone(3P)$ mass,
\begin{equation*}
m(\chibone(3P))= 10509.0^{+5.0}_{-2.6}\stat ^{+2.5}_{-2.4}\syst \mevcc,
\end{equation*}
is derived from the radiative transition to the $\OneS$ meson, where the $\chib(3P)$ is observed with a statistical significance of 6.0$\sigma$.
Another measurement,
\begin{equation*}
m(\chibone(3P))= 10518.5^{+1.9}_{-1.3}\stat \pm0.6\syst \mevcc,
\end{equation*}
is derived from the radiative transition to the $\TwoS$ transition, where evidence is found for the $\chib(3P)$ with a statistical significance of 3.6$\sigma$.
The systematic uncertainty related to  $r_{12}$  is largely uncorrelated between the $\TwoS$ and $\OneS$ channels as the branching fractions
of $\chi_{bi}$  to  final states involving $\OneS$ and to $\TwoS$ mesons can be different. 
By treating the systematic uncertainties related to the mass splitting and to the mass bias as fully correlated and all other uncertainties 
as uncorrelated, the two results for the $\chibone(3P)$ mass differ 
by $9.3^{+3.2}_{-5.2}\stat\pm 2.0\syst$~\mevcc.
A combined fit is performed leading to
\begin{equation*}
m(\chibone(3P))= 10515.7^{+2.2}_{-3.9}\stat ^{+1.5}_{-2.1}\syst \mevcc.
\end{equation*}
In these measurements, the relative rate  of $\chibtwo$ to $\chibone$, is  assumed to be $r_{12}=0.42$ for the two transitions.
The $\chibone(3P)$ mass result exhibits a linear dependence on the assumed fraction of $\chibone$ decays and varies from $10517.6$ to $10515.2$ when
the $\chibtwo/\chibone$ yield ratio changes from zero to 0.5.
This result is compatible with and significantly more precise than that reported by the ATLAS experiment, 
$m(\chib(3P))= 10530\pm5\stat \pm 9\syst \mevcc$ for $r_{12}=1$ and $\Delta m_{12}=12$\mevcc, where $m(\chib(3P))$ is the average mass 
of $\chibone$ and $\chibtwo$ states~\cite{ATLASchib}. 
The LHCb result is also compatible with the D0 measurement, $m(\chib(3P))= 10551\pm14\stat \pm 17\syst \mevcc$~\cite{D0chib}.

\begin{figure}[tb]
  \begin{center}
    \begin{tabular}{cc}
      \hspace{-.9cm}    \includegraphics[width=0.55\linewidth]{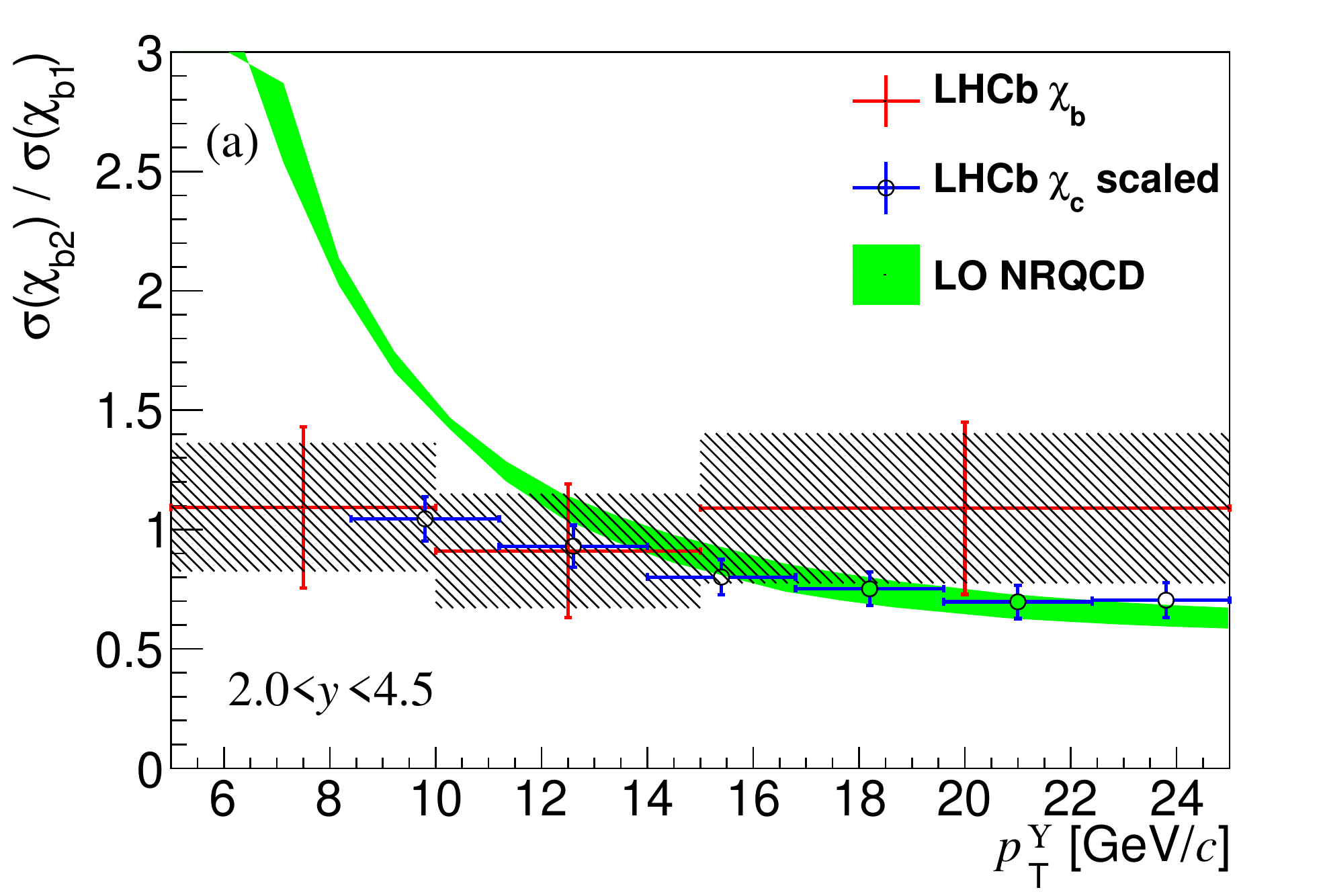} &
      \hspace{-.9cm}    \includegraphics[width=0.55\linewidth]{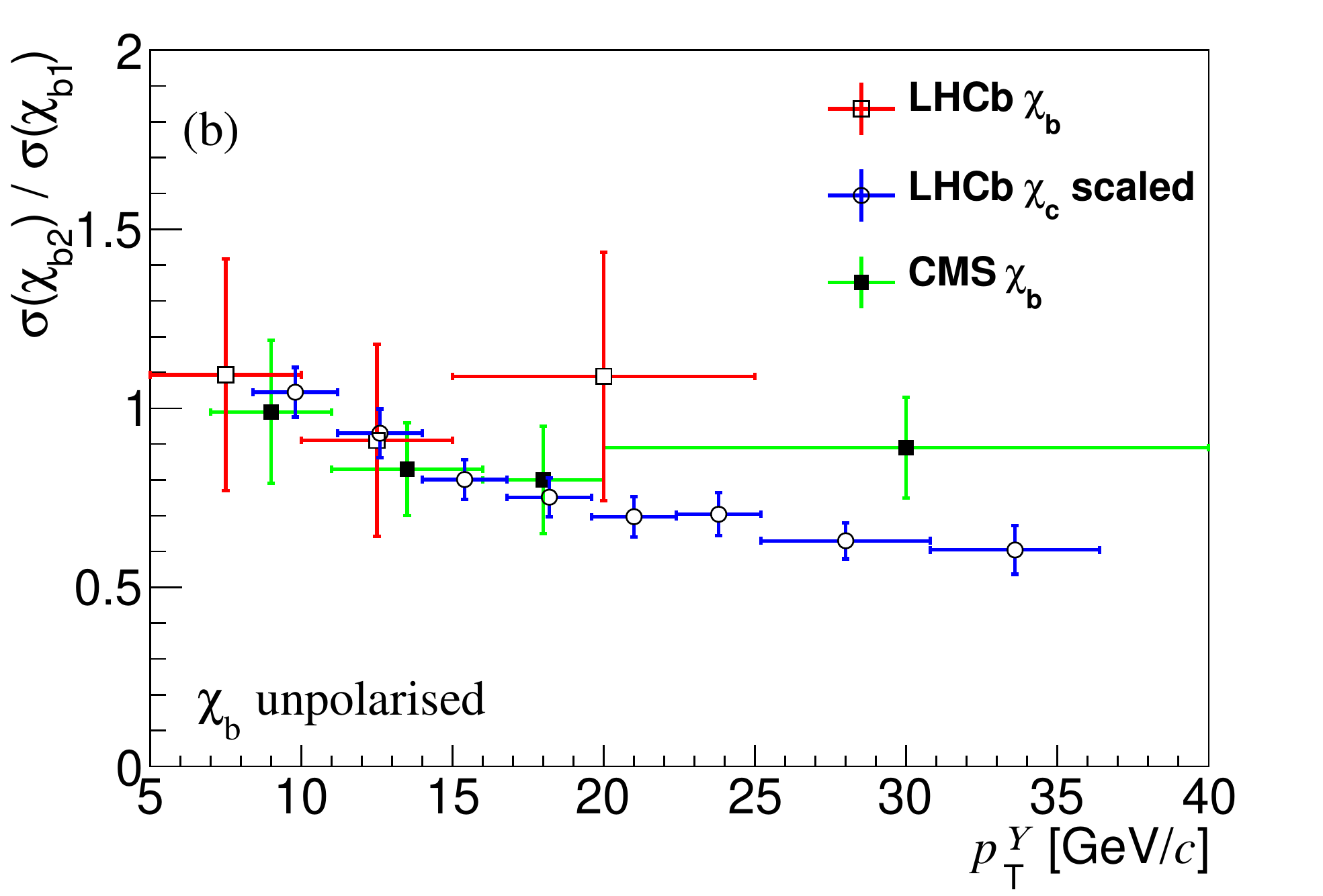} \\
    \vspace*{-1.0cm}
    \end{tabular}
  \end{center}
  \caption{\small Relative production cross-sections of $\chi_{b1}$ to $\chi_{b2}$ mesons as a function of $\ptupsilon$. 
Panel (a) shows the comparison of this measurement 
(the  hatched rectangles show the statistical uncertainties and the red crosses the total experimental uncertainty)
to the LO NRQCD prediction~\cite{Likhoded_chib} (green band),
and to the LHCb $\chic$ result (blue crosses), where the \pt axis has been scaled by $m(\chib)/m(\chic)=2.8$. 
Panel (b) compares this measurement (empty squares) to CMS results~\cite{CMSchib} (filled squares) and to the scaled
LHCb $\chic$ results (empty circles). The error bars are the total experimental uncertainties and do not include the uncertainties on 
the branching fractions. 
  \label{fig:sigmaratio}}
\end{figure}

The ratio of the $\chibtwo$ to $\chibone$ production cross-sections is measured in three \ptupsilon ranges using Eq.~(\ref{eq:rxs}). 
The results are given in Table~\ref{tbl:sigmaratio}.
Figure~\ref{fig:sigmaratio} (a) shows a comparison of the measured values with LO NRQCD predictions from Ref.~\cite{Likhoded_chib}.
The common systematic uncertainty ($9.0\%$) due to the branching fraction of $\chib\to\OneS\gamma$ is not shown.
Theory predicts the $\chic$ and $\chib$ ratio of production cross-section to be the same when the $\chic$ \pt value is scaled by the ratio of the 
$\chib$ and $\chic$ masses~\cite{Likhoded_chib}. 
As the \chib (\chic) and $\PUpsilon$ ($\jpsi$) \pt are strongly correlated, this is assumed to be valid when
replacing the \chib (\chic) by the $\PUpsilon$ ($\jpsi$) \pt.
The measurement obtained by LHCb for the $\chic$ production ratio~\cite{LHCb-PAPER-2013-028} with the \pt axis scaled accordingly is also 
shown for comparison.
The $\chib$ results are in good agreement with the scaled $\chic$ results. 
These results are not precise enough to establish the deviation from unity predicted by theory at low \pt, but the agreement is better with a flat dependence. 
Our results are also in agreement with the CMS results~\cite{CMSchib} as shown on Fig.~\ref{fig:sigmaratio} (b).
\begin{table}
\begin{center}
\caption{\small Relative production cross section of $\chi_{b1}$ to $\chi_{b2}$ mesons for the $1P$ state for each \ptupsilon bin. 
The first uncertainty is statistical, the second 
is the systematic uncertainty and the third is due to the uncertainty on the branching fractions.}
\label{tbl:sigmaratio}
{\renewcommand{\arraystretch}{1.4}
\begin{tabular}{l c}
\ptupsilon bin (\gevc)       & $\sigma(\chi_{b2})/\sigma(\chi_{b1})$ \\
\hline

5--10 & $1.09\pm0.27\stat ^{+0.11}_{-0.18}\syst \pm 0.10\,(\BR)$ \\ 
10--15 & $0.91\pm0.24\stat ^{+0.10}_{-0.12}\syst \pm 0.08\,(\BR)$ \\ 
15--25 & $1.09\pm0.31\stat ^{+0.14}_{-0.15}\syst \pm0.10\,(\BR)$  \\ 
\end{tabular}
}
\end{center}
\end{table}
%
\section{Conclusion}
The radiative decays of $\chib$ mesons to $\PUpsilon$ mesons are reconstructed with photons converting in the detector material.
Owing to the good energy resolution obtained with converted photons, the $\chib(1P)$ states are separated and the mass splitting between
the $\chibone(1P)$ and $\chibtwo(1P)$ is measured.
The $\chib(3P)$ mass is measured using its radiative decays to the $\OneS$ and $\TwoS$ mesons yielding,
\begin{equation*}
m(\chibone(3P))= 10515.7^{+2.2}_{-3.9}\stat ^{+1.5}_{-2.1}\syst \mevcc.
\end{equation*}
This result is compatible with the measurement performed by LHCb with the radiative decays to the $\ThreeS$ meson that uses
non-converted photons~\cite{LHCb-PAPER-2014-031}, $m(\chibone(3P))= 10511.3\pm1.7\stat\pm2.5\syst$~\mevcc. 
Since the photon reconstruction is based on different 
subdetectors, the experimental systematic uncertainties are uncorrelated, while the uncertainty related to the model used for 
summing the $J=1$ and $J=2$ contributions
(parametrised with the mass splitting $\Delta m_{12}$ and the relative rates $r_{12}$) are fully correlated. The
combined value is
\begin{equation*}
m(\chibone(3P))= 10512.1 \pm 2.1\expe \pm 0.9\model \mevcc,
\end{equation*}
where the first uncertainty is experimental
(statistical and systematic) and the second accounts for varying $\Delta m_{12}$ from 9.0 to 12.0~\mevcc and $r_{12}$ by $\pm 30\%$.
This result is in agreement with the theoretical prediction of Ref.~\cite{Rosner-Kwong}, $m(\chibone(3P))=10516~\mevcc$.

The first measurement of the relative ratio of $\chi_{b1}$ to $\chi_{b2}$ cross-sections is performed for the $\chib(1P)$ 
state in the rapidity range $2.0<y<4.5$ for \ptupsilon from 5 to 25~\gevc. 
The results agree with CMS results~\cite{CMSchib} and with theory expectation based on LHCb $\chic$ measurements~\cite{LHCb-PAPER-2013-028}.
The data indicate a deviation from the rise predicted by the LO NRQCD model at low \pt and show a better agreement with a flat dependence.

\section*{Acknowledgements}
\noindent 
We thank A. Luchinsky and A. Likhoded for providing the 
LO NRQCD predictions.
We express our gratitude to our colleagues in the CERN
accelerator departments for the excellent performance of the LHC. We
thank the technical and administrative staff at the LHCb
institutes. We acknowledge support from CERN and from the national
agencies: CAPES, CNPq, FAPERJ and FINEP (Brazil); NSFC (China);
CNRS/IN2P3 (France); BMBF, DFG, HGF and MPG (Germany); SFI (Ireland); INFN (Italy); 
FOM and NWO (The Netherlands); MNiSW and NCN (Poland); MEN/IFA (Romania); 
MinES and FANO (Russia); MinECo (Spain); SNSF and SER (Switzerland); 
NASU (Ukraine); STFC (United Kingdom); NSF (USA).
The Tier1 computing centres are supported by IN2P3 (France), KIT and BMBF 
(Germany), INFN (Italy), NWO and SURF (The Netherlands), PIC (Spain), GridPP 
(United Kingdom).
We are indebted to the communities behind the multiple open 
source software packages on which we depend. We are also thankful for the 
computing resources and the access to software R\&D tools provided by Yandex LLC (Russia).
Individual groups or members have received support from 
EPLANET, Marie Sk\l{}odowska-Curie Actions and ERC (European Union), 
Conseil g\'{e}n\'{e}ral de Haute-Savoie, Labex ENIGMASS and OCEVU, 
R\'{e}gion Auvergne (France), RFBR (Russia), XuntaGal and GENCAT (Spain), Royal Society and Royal
Commission for the Exhibition of 1851 (United Kingdom).

\addcontentsline{toc}{section}{References}
\setboolean{inbibliography}{true}
\bibliographystyle{LHCb}
\bibliography{main}

\ifx\mcitethebibliography\mciteundefinedmacro
\PackageError{LHCb.bst}{mciteplus.sty has not been loaded}
{This bibstyle requires the use of the mciteplus package.}\fi
\providecommand{\href}[2]{#2}
\begin{mcitethebibliography}{10}
\mciteSetBstSublistMode{n}
\mciteSetBstMaxWidthForm{subitem}{\alph{mcitesubitemcount})}
\mciteSetBstSublistLabelBeginEnd{\mcitemaxwidthsubitemform\space}
{\relax}{\relax}

\bibitem{CSMBaier}
R.~Baier and R.~R{\"u}ckl, \ifthenelse{\boolean{articletitles}}{{\it Hadronic
  collisions: a quarkonium factory},
  }{}\href{http://dx.doi.org/10.1007/BF01572254}{Z.\ Phys.\  {\bf C19} (1983)
  251}\relax
\mciteBstWouldAddEndPuncttrue
\mciteSetBstMidEndSepPunct{\mcitedefaultmidpunct}
{\mcitedefaultendpunct}{\mcitedefaultseppunct}\relax
\EndOfBibitem
\bibitem{CSMLikhoded}
V.~Kartvelishvili, A.~Likhoded, and S.~Slabospitsky,
  \ifthenelse{\boolean{articletitles}}{{\it {$D$ meson and $\psi$ meson
  production in hadronic interactions}}, }{}Sov.\ J.\ Nucl.\ Phys.\  {\bf 28}
  (1978) 678\relax
\mciteBstWouldAddEndPuncttrue
\mciteSetBstMidEndSepPunct{\mcitedefaultmidpunct}
{\mcitedefaultendpunct}{\mcitedefaultseppunct}\relax
\EndOfBibitem
\bibitem{CSMBerger}
E.~L. Berger and D.~Jones, \ifthenelse{\boolean{articletitles}}{{\it {Inelastic
  photoproduction of $\jpsi$ and $\PUpsilon$ by gluons}},
  }{}\href{http://dx.doi.org/10.1103/PhysRevD.23.1521}{Phys.\ Rev.\  {\bf D23}
  (1981) 1521}\relax
\mciteBstWouldAddEndPuncttrue
\mciteSetBstMidEndSepPunct{\mcitedefaultmidpunct}
{\mcitedefaultendpunct}{\mcitedefaultseppunct}\relax
\EndOfBibitem
\bibitem{NRQCD0}
G.~T. Bodwin, E.~Braaten, and G.~Lepage,
  \ifthenelse{\boolean{articletitles}}{{\it {Rigorous QCD analysis of inclusive
  annihilation and production of heavy quarkonium}},
  }{}\href{http://dx.doi.org/10.1103/PhysRevD.51.1125}{Phys.\ Rev.\  {\bf D51}
  (1995) 1125}\relax
\mciteBstWouldAddEndPuncttrue
\mciteSetBstMidEndSepPunct{\mcitedefaultmidpunct}
{\mcitedefaultendpunct}{\mcitedefaultseppunct}\relax
\EndOfBibitem
\bibitem{NRQCD0err}
G.~T. Bodwin, E.~Braaten, and G.~Lepage,
  \ifthenelse{\boolean{articletitles}}{{\it {Erratum: Rigorous QCD analysis of
  inclusive annihilation and production of heavy quarkonium}},
  }{}\href{http://dx.doi.org/http://dx.doi.org/10.1103/PhysRevD.55.5853}{Phys.\
  Rev.\  {\bf D55} (1997) 5853}\relax
\mciteBstWouldAddEndPuncttrue
\mciteSetBstMidEndSepPunct{\mcitedefaultmidpunct}
{\mcitedefaultendpunct}{\mcitedefaultseppunct}\relax
\EndOfBibitem
\bibitem{NRQCD}
Y.~Q. Ma, K.~Wang, and K.~T. Chao, \ifthenelse{\boolean{articletitles}}{{\it
  {\rm QCD} radiative corrections to $\chi_{cJ}$ production at hadron
  colliders}, }{}\href{http://dx.doi.org/10.1103/PhysRevD.83.111503}{Phys.\
  Rev.\  {\bf D83} (2011) 111503}, \href{http://arxiv.org/abs/1002.3987}{{\tt
  arXiv:1002.3987}}\relax
\mciteBstWouldAddEndPuncttrue
\mciteSetBstMidEndSepPunct{\mcitedefaultmidpunct}
{\mcitedefaultendpunct}{\mcitedefaultseppunct}\relax
\EndOfBibitem
\bibitem{Lansberg}
J.-P. Lansberg, \ifthenelse{\boolean{articletitles}}{{\it On the mechanisms of
  heavy-quarkonium hadroproduction},
  }{}\href{http://dx.doi.org/10.1140/epjc/s10052-008-0826-9}{Eur.\ Phys.\ J.\
  {\bf C61} (2009) 693}, \href{http://arxiv.org/abs/0811.4005}{{\tt
  arXiv:0811.4005}}\relax
\mciteBstWouldAddEndPuncttrue
\mciteSetBstMidEndSepPunct{\mcitedefaultmidpunct}
{\mcitedefaultendpunct}{\mcitedefaultseppunct}\relax
\EndOfBibitem
\bibitem{Tramontano}
J.~Campbell, F.~Maltoni, and F.~Tramontano,
  \ifthenelse{\boolean{articletitles}}{{\it {QCD corrections to $\jpsi$ and
  $\Upsilon$ production at hadron colliders}},
  }{}\href{http://dx.doi.org/10.1103/PhysRevLett.98.252002}{Phys.\ Rev.\ Lett.\
   {\bf 98} (2007) 252002}, \href{http://arxiv.org/abs/hep-ph/0703113}{{\tt
  arXiv:hep-ph/0703113}}\relax
\mciteBstWouldAddEndPuncttrue
\mciteSetBstMidEndSepPunct{\mcitedefaultmidpunct}
{\mcitedefaultendpunct}{\mcitedefaultseppunct}\relax
\EndOfBibitem
\bibitem{Likhoded_chib}
A.~Likhoded, A.~Luchinsky, and S.~Poslavsky,
  \ifthenelse{\boolean{articletitles}}{{\it Production of $\chi_b$-mesons at
  \lhc}, }{}\href{http://dx.doi.org/10.1103/PhysRevD.86.074027}{Phys.\ Rev.\
  {\bf D86} (2012) 074027}, \href{http://arxiv.org/abs/1203.4893}{{\tt
  arXiv:1203.4893}}\relax
\mciteBstWouldAddEndPuncttrue
\mciteSetBstMidEndSepPunct{\mcitedefaultmidpunct}
{\mcitedefaultendpunct}{\mcitedefaultseppunct}\relax
\EndOfBibitem
\bibitem{WA11chic}
WA11 collaboration, Y.~Lemoigne {\em et~al.},
  \ifthenelse{\boolean{articletitles}}{{\it Measurement of hadronic production
  of the $\chi_1^{++}(3507)$ and the $\chi_2^{++}(3553)$ through their
  radiative decay to \jpsi},
  }{}\href{http://dx.doi.org/10.1016/0370-2693(82)90795-X}{Phys.\ Lett.\  {\bf
  B113} (1982) 509}\relax
\mciteBstWouldAddEndPuncttrue
\mciteSetBstMidEndSepPunct{\mcitedefaultmidpunct}
{\mcitedefaultendpunct}{\mcitedefaultseppunct}\relax
\EndOfBibitem
\bibitem{HERABchic}
HERA-B collaboration, I.~Abt {\em et~al.},
  \ifthenelse{\boolean{articletitles}}{{\it Production of the charmonium states
  \chicone and \chictwo in proton nucleus interactions at $\sqs=41.6$ \gev},
  }{}\href{http://dx.doi.org/10.1103/PhysRevD.79.012001}{Phys.\ Rev.\  {\bf
  D79} (2009) 012001}, \href{http://arxiv.org/abs/0807.2167}{{\tt
  arXiv:0807.2167}}\relax
\mciteBstWouldAddEndPuncttrue
\mciteSetBstMidEndSepPunct{\mcitedefaultmidpunct}
{\mcitedefaultendpunct}{\mcitedefaultseppunct}\relax
\EndOfBibitem
\bibitem{CDFchic}
CDF collaboration, A.~Abulencia {\em et~al.},
  \ifthenelse{\boolean{articletitles}}{{\it {Measurement of
  ${\sigma_{\chictwo}\BF(\chictwo\to\jpsi\gamma)/\sigma_{\chicone}\BF(\chicone
  \to \jpsi\gamma)}$ in ${\ensuremath{\Pp}}\antiproton$ collisions at
  $\sqs=1.96$ \tev}},
  }{}\href{http://dx.doi.org/10.1103/PhysRevLett.98.232001}{Phys.\ Rev.\ Lett.\
   {\bf 98} (2007) 232001}, \href{http://arxiv.org/abs/hep-ph/0703028}{{\tt
  arXiv:hep-ph/0703028}}\relax
\mciteBstWouldAddEndPuncttrue
\mciteSetBstMidEndSepPunct{\mcitedefaultmidpunct}
{\mcitedefaultendpunct}{\mcitedefaultseppunct}\relax
\EndOfBibitem
\bibitem{CMSchic}
CMS collaboration, S.~Chatrchyan {\em et~al.},
  \ifthenelse{\boolean{articletitles}}{{\it {Measurement of the relative prompt
  production rate of $\chictwo$ and $\chicone$ in pp collisions at $\sqs=7$
  \tev}}, }{}\href{http://dx.doi.org/10.1140/epjc/s10052-012-2251-3}{Eur.\
  Phys.\ J.\  {\bf C72} (2012) 2251},
  \href{http://arxiv.org/abs/1210.0875}{{\tt arXiv:1210.0875}}\relax
\mciteBstWouldAddEndPuncttrue
\mciteSetBstMidEndSepPunct{\mcitedefaultmidpunct}
{\mcitedefaultendpunct}{\mcitedefaultseppunct}\relax
\EndOfBibitem
\bibitem{LHCb-PAPER-2013-028}
LHCb collaboration, R.~Aaij {\em et~al.},
  \ifthenelse{\boolean{articletitles}}{{\it {Measurement of the relative rate
  of prompt $\chi_{c0}$, $\chi_{c1}$ and $\chi_{c2}$ production at $\sqrt{s}=7
  \tev$}}, }{}\href{http://dx.doi.org/10.1007/JHEP10(2013)115}{JHEP {\bf 10}
  (2013) 115}, \href{http://arxiv.org/abs/1307.4285}{{\tt
  arXiv:1307.4285}}\relax
\mciteBstWouldAddEndPuncttrue
\mciteSetBstMidEndSepPunct{\mcitedefaultmidpunct}
{\mcitedefaultendpunct}{\mcitedefaultseppunct}\relax
\EndOfBibitem
\bibitem{ATLASchib}
ATLAS collaboration, G.~Aad {\em et~al.},
  \ifthenelse{\boolean{articletitles}}{{\it {Observation of a new $\chib$ state
  in radiative transistions to $\PUpsilon$(1S) and $\PUpsilon$(2S) at ATLAS}},
  }{}\href{http://dx.doi.org/10.1103/PhysRevLett.108.152001}{Phys.\ Rev.\ Lett
  {\bf 108} (2012) 152001}, \href{http://arxiv.org/abs/1112.5154}{{\tt
  arXiv:1112.5154}}\relax
\mciteBstWouldAddEndPuncttrue
\mciteSetBstMidEndSepPunct{\mcitedefaultmidpunct}
{\mcitedefaultendpunct}{\mcitedefaultseppunct}\relax
\EndOfBibitem
\bibitem{D0chib}
D0 collaboration, V.~Abazov {\em et~al.},
  \ifthenelse{\boolean{articletitles}}{{\it {Observation of a narrow mass state
  decaying into $\Upsilon$(1S) + $\gamma$ in $\antiproton{\ensuremath{\Pp}}$
  collisions at $\sqrt{s}=1.96 \tev$}},
  }{}\href{http://dx.doi.org/10.1103/PhysRevD.86.031103}{Phys.\ Rev.\  {\bf
  D86} (2012) 031103}, \href{http://arxiv.org/abs/1203.6034}{{\tt
  arXiv:1203.6034}}\relax
\mciteBstWouldAddEndPuncttrue
\mciteSetBstMidEndSepPunct{\mcitedefaultmidpunct}
{\mcitedefaultendpunct}{\mcitedefaultseppunct}\relax
\EndOfBibitem
\bibitem{thchibmass}
L.~Motyka and K.~Zalewski, \ifthenelse{\boolean{articletitles}}{{\it Mass
  spectra and leptonic decay widths of heavy quarkonia},
  }{}\href{http://dx.doi.org/10.1007/s100529800743}{Eur.\ Phys.\ J.\  {\bf C4}
  (1998) 107}\relax
\mciteBstWouldAddEndPuncttrue
\mciteSetBstMidEndSepPunct{\mcitedefaultmidpunct}
{\mcitedefaultendpunct}{\mcitedefaultseppunct}\relax
\EndOfBibitem
\bibitem{Rosner-Kwong}
W.~Kwong and J.~Rosner, \ifthenelse{\boolean{articletitles}}{{\it {D-wave
  quarkonium levels of the $\Upsilon$ family}},
  }{}\href{http://dx.doi.org/10.1103/PhysRevD.38.279}{Phys.\ Rev.\  {\bf D38}
  (1988) 279}\relax
\mciteBstWouldAddEndPuncttrue
\mciteSetBstMidEndSepPunct{\mcitedefaultmidpunct}
{\mcitedefaultendpunct}{\mcitedefaultseppunct}\relax
\EndOfBibitem
\bibitem{Ferretti-Galata}
J.~Ferretti and G.~Galat\`a, \ifthenelse{\boolean{articletitles}}{{\it {Quark
  structure of the $X(3872)$ and $\chib(3P)$ resonances}},
  }{}\href{http://arxiv.org/abs/1401.4431}{{\tt arXiv:1401.4431}}\relax
\mciteBstWouldAddEndPuncttrue
\mciteSetBstMidEndSepPunct{\mcitedefaultmidpunct}
{\mcitedefaultendpunct}{\mcitedefaultseppunct}\relax
\EndOfBibitem
\bibitem{PDG2012}
Particle Data Group, J.~Beringer {\em et~al.},
  \ifthenelse{\boolean{articletitles}}{{\it {\href{http://pdg.lbl.gov/}{Review
  of particle physics}}},
  }{}\href{http://dx.doi.org/10.1103/PhysRevD.86.010001}{Phys.\ Rev.\  {\bf
  D86} (2012) 010001}, {and 2013 partial update for the 2014 edition}\relax
\mciteBstWouldAddEndPuncttrue
\mciteSetBstMidEndSepPunct{\mcitedefaultmidpunct}
{\mcitedefaultendpunct}{\mcitedefaultseppunct}\relax
\EndOfBibitem
\bibitem{Alves:2008zz}
LHCb collaboration, A.~A. Alves~Jr. {\em et~al.},
  \ifthenelse{\boolean{articletitles}}{{\it {The \lhcb detector at the LHC}},
  }{}\href{http://dx.doi.org/10.1088/1748-0221/3/08/S08005}{JINST {\bf 3}
  (2008) S08005}\relax
\mciteBstWouldAddEndPuncttrue
\mciteSetBstMidEndSepPunct{\mcitedefaultmidpunct}
{\mcitedefaultendpunct}{\mcitedefaultseppunct}\relax
\EndOfBibitem
\bibitem{Sjostrand:2006za}
T.~Sj\"{o}strand, S.~Mrenna, and P.~Skands,
  \ifthenelse{\boolean{articletitles}}{{\it {PYTHIA 6.4 physics and manual}},
  }{}\href{http://dx.doi.org/10.1088/1126-6708/2006/05/026}{JHEP {\bf 05}
  (2006) 026}, \href{http://arxiv.org/abs/hep-ph/0603175}{{\tt
  arXiv:hep-ph/0603175}}\relax
\mciteBstWouldAddEndPuncttrue
\mciteSetBstMidEndSepPunct{\mcitedefaultmidpunct}
{\mcitedefaultendpunct}{\mcitedefaultseppunct}\relax
\EndOfBibitem
\bibitem{Sjostrand:2007gs}
T.~Sj\"{o}strand, S.~Mrenna, and P.~Skands,
  \ifthenelse{\boolean{articletitles}}{{\it {A brief introduction to PYTHIA
  8.1}}, }{}\href{http://dx.doi.org/10.1016/j.cpc.2008.01.036}{Comput.\ Phys.\
  Commun.\  {\bf 178} (2008) 852}, \href{http://arxiv.org/abs/0710.3820}{{\tt
  arXiv:0710.3820}}\relax
\mciteBstWouldAddEndPuncttrue
\mciteSetBstMidEndSepPunct{\mcitedefaultmidpunct}
{\mcitedefaultendpunct}{\mcitedefaultseppunct}\relax
\EndOfBibitem
\bibitem{LHCb-PROC-2010-056}
I.~Belyaev {\em et~al.}, \ifthenelse{\boolean{articletitles}}{{\it {Handling of
  the generation of primary events in \gauss, the \lhcb simulation framework}},
  }{}\href{http://dx.doi.org/10.1109/NSSMIC.2010.5873949}{Nuclear Science
  Symposium Conference Record (NSS/MIC) {\bf IEEE} (2010) 1155}\relax
\mciteBstWouldAddEndPuncttrue
\mciteSetBstMidEndSepPunct{\mcitedefaultmidpunct}
{\mcitedefaultendpunct}{\mcitedefaultseppunct}\relax
\EndOfBibitem
\bibitem{Lange:2001uf}
D.~J. Lange, \ifthenelse{\boolean{articletitles}}{{\it {The EvtGen particle
  decay simulation package}},
  }{}\href{http://dx.doi.org/10.1016/S0168-9002(01)00089-4}{Nucl.\ Instrum.\
  Meth.\  {\bf A462} (2001) 152}\relax
\mciteBstWouldAddEndPuncttrue
\mciteSetBstMidEndSepPunct{\mcitedefaultmidpunct}
{\mcitedefaultendpunct}{\mcitedefaultseppunct}\relax
\EndOfBibitem
\bibitem{Golonka:2005pn}
P.~Golonka and Z.~Was, \ifthenelse{\boolean{articletitles}}{{\it {PHOTOS Monte
  Carlo: A precision tool for QED corrections in $Z$ and $W$ decays}},
  }{}\href{http://dx.doi.org/10.1140/epjc/s2005-02396-4}{Eur.\ Phys.\ J.\  {\bf
  C45} (2006) 97}, \href{http://arxiv.org/abs/hep-ph/0506026}{{\tt
  arXiv:hep-ph/0506026}}\relax
\mciteBstWouldAddEndPuncttrue
\mciteSetBstMidEndSepPunct{\mcitedefaultmidpunct}
{\mcitedefaultendpunct}{\mcitedefaultseppunct}\relax
\EndOfBibitem
\bibitem{Allison:2006ve}
Geant4 collaboration, J.~Allison {\em et~al.},
  \ifthenelse{\boolean{articletitles}}{{\it {Geant4 developments and
  applications}}, }{}\href{http://dx.doi.org/10.1109/TNS.2006.869826}{IEEE
  Trans.\ Nucl.\ Sci.\  {\bf 53} (2006) 270}\relax
\mciteBstWouldAddEndPuncttrue
\mciteSetBstMidEndSepPunct{\mcitedefaultmidpunct}
{\mcitedefaultendpunct}{\mcitedefaultseppunct}\relax
\EndOfBibitem
\bibitem{Agostinelli:2002hh}
Geant4 collaboration, S.~Agostinelli {\em et~al.},
  \ifthenelse{\boolean{articletitles}}{{\it {Geant4: A simulation toolkit}},
  }{}\href{http://dx.doi.org/10.1016/S0168-9002(03)01368-8}{Nucl.\ Instrum.\
  Meth.\  {\bf A506} (2003) 250}\relax
\mciteBstWouldAddEndPuncttrue
\mciteSetBstMidEndSepPunct{\mcitedefaultmidpunct}
{\mcitedefaultendpunct}{\mcitedefaultseppunct}\relax
\EndOfBibitem
\bibitem{LHCb-PROC-2011-006}
M.~Clemencic {\em et~al.}, \ifthenelse{\boolean{articletitles}}{{\it {The \lhcb
  simulation application, \gauss: design, evolution and experience}},
  }{}\href{http://dx.doi.org/10.1088/1742-6596/331/3/032023}{{J.\ Phys.\ \!\!:
  Conf.\ Ser.\ } {\bf 331} (2011) 032023}\relax
\mciteBstWouldAddEndPuncttrue
\mciteSetBstMidEndSepPunct{\mcitedefaultmidpunct}
{\mcitedefaultendpunct}{\mcitedefaultseppunct}\relax
\EndOfBibitem
\bibitem{Skwarnicki:1986xj}
T.~Skwarnicki, {\em {A study of the radiative cascade transitions between the
  Upsilon-prime and Upsilon resonances}}, PhD thesis, Institute of Nuclear
  Physics, Krakow, 1986,
  {\href{http://inspirehep.net/record/230779/files/230779.pdf}{DESY-F31-86-02}%
}\relax
\mciteBstWouldAddEndPuncttrue
\mciteSetBstMidEndSepPunct{\mcitedefaultmidpunct}
{\mcitedefaultendpunct}{\mcitedefaultseppunct}\relax
\EndOfBibitem
\bibitem{CMSchib}
CMS collaboration, V.~Khachatryan {\em et~al.},
  \ifthenelse{\boolean{articletitles}}{{\it {Measurement of the production
  cross section ratio $\sigma(\chibtwo(1P))/\sigma(\chibone(1P))$ in pp
  collisions at $\sqrt{s}=8 \tev$}},
  }{}\href{http://arxiv.org/abs/1409.5761}{{\tt arXiv:1409.5761}}\relax
\mciteBstWouldAddEndPuncttrue
\mciteSetBstMidEndSepPunct{\mcitedefaultmidpunct}
{\mcitedefaultendpunct}{\mcitedefaultseppunct}\relax
\EndOfBibitem
\bibitem{LHCb-PAPER-2014-031}
LHCb collaboration, R.~Aaij {\em et~al.},
  \ifthenelse{\boolean{articletitles}}{{\it {Study of $\chi_b$ meson production
  in $pp$ collisions at $\sqrt{s}$=7 and 8 TeV and observation of the decay
  $\chi_b \to \Upsilon(3S)\gamma$}},
  }{}\href{http://arxiv.org/abs/1407.7734}{{\tt arXiv:1407.7734}}, {submited to
  Eur. Phys. J. C}\relax
\mciteBstWouldAddEndPuncttrue
\mciteSetBstMidEndSepPunct{\mcitedefaultmidpunct}
{\mcitedefaultendpunct}{\mcitedefaultseppunct}\relax
\EndOfBibitem
\end{mcitethebibliography}
\newpage
\centerline{\large\bf LHCb collaboration}
\begin{flushleft}
\small
R.~Aaij$^{41}$, 
B.~Adeva$^{37}$, 
M.~Adinolfi$^{46}$, 
A.~Affolder$^{52}$, 
Z.~Ajaltouni$^{5}$, 
S.~Akar$^{6}$, 
J.~Albrecht$^{9}$, 
F.~Alessio$^{38}$, 
M.~Alexander$^{51}$, 
S.~Ali$^{41}$
% ...

{\footnotesize \it
% ...
$ ^{5}$Clermont Universit\'{e}, Universit\'{e} Blaise Pascal, CNRS/IN2P3, LPC, Clermont-Ferrand, France\\
$ ^{6}$CPPM, Aix-Marseille Universit\'{e}, CNRS/IN2P3, Marseille, France\\
% ...
$ ^{9}$Fakult\"{a}t Physik, Technische Universit\"{a}t Dortmund, Dortmund, Germany\\
% ...
$ ^{37}$Universidad de Santiago de Compostela, Santiago de Compostela, Spain\\
$ ^{38}$European Organization for Nuclear Research (CERN), Geneva, Switzerland\\
% ...
$ ^{41}$Nikhef National Institute for Subatomic Physics, Amsterdam, The Netherlands\\
% ...
$ ^{46}$H.H. Wills Physics Laboratory, University of Bristol, Bristol, United Kingdom\\
% ...
$ ^{51}$School of Physics and Astronomy, University of Glasgow, Glasgow, United Kingdom\\
$ ^{52}$Oliver Lodge Laboratory, University of Liverpool, Liverpool, United Kingdom\\
% ...
}
\end{flushleft}

\end{document}